\documentclass[a4paper,11pt]{article}
\usepackage{aaskaiid}
\usepackage{hyperref}
\usepackage{xspace}
\usepackage{comment}
\usepackage{wrapfig}
\usepackage{orcidlink}

\newcommand{\arcsec}{\ensuremath{^{\prime\prime}}\xspace}
\newcommand{\farcs}{.\!\!^{\prime\prime}}
\newcommand{\arcmin}{\ensuremath{^{\prime}}\xspace}

\title{Chemical Complexity in the Early Stages of Star Formation in the SKAO Era}
\ShortTitle{Chemical Complexity}

\author[1]{Eleonora Bianchi\orcidlink{0000-0001-9249-7082}}
\ShortName{Bianchi et al.} 
\author[2]{Mathilde Bouvier\orcidlink{0000-0003-0167-0746}}
\author[1]{Claudio Codella\orcidlink{0000-0003-1514-3074}}
\author[3]{Laura Colzi\orcidlink{0000-0001-8064-6394}}
\author[4]{Audrey Coutens\orcidlink{0000-0003-1805-3920}}
\author[5]{Marta De Simone\orcidlink{0000-0001-5659-0140}}
\author[6]{Joan Enrique Romero\orcidlink{0000-0002-2147-7735}}
\author[7]{Gisela Esplugues\orcidlink{0000-0002-4292-4127}}
\author[1,15]{Francesco Fontani\orcidlink{0000-0003-0348-3418}}
\author[8]{Antonio Garufi\orcidlink{0000-0002-4266-0643}}
\author[1]{Lisa Giani\orcidlink{0000-0002-7327-9132}}
\author[9,10]{Arshia Maria Jacob\orcidlink{0000-0001-7838-3425}}
\author[3]{Izaskun Jiménez-Serra\orcidlink{0000-0003-4493-8714}}
\author[1]{Marco Padovani\orcidlink{0000-0003-2303-0096}}
\author[1]{Linda Podio\orcidlink{0000-0003-2733-5372}}
\author[11]{Albert Rimola\orcidlink{0000-0002-9637-4554}}
\author[7]{Pablo Rivière Marichalar\orcidlink{0000-0003-0969-8137}}
\author[1]{Giovanni Sabatini\orcidlink{0000-0002-6428-9806}}
\author[12]{Andrea Socci\orcidlink{0009-0008-0541-572X}}
\author[13]{Riccardo Giovanni Urso\orcidlink{0000-0001-6926-1434}}

\author[14]{Tyler L. Bourke\orcidlink{0000-0001-7491-0048}}
\author[15]{Gemma Busquet\orcidlink{0000-0002-2189-6278}}
\author[16]{Paola Caselli\orcidlink{0000-0003-1481-7911}}
\author[17]{Cecilia Ceccarelli\orcidlink{0000-0001-9664-6292}}
\author[18]{Tomoya Hirota\orcidlink{0000-0003-1659-095X}}
\author[19]{John D. Ilee\orcidlink{0000-0003-1008-1142}}
\author[16]{Valerio Lattanzi\orcidlink{0000-0001-9819-1658}}
\author[1]{Manuela Lippi\orcidlink{0000-0001-9185-878X}}
\author[17]{Ana López-Sepulcre\orcidlink{0000-0002-6729-3640}}
\author[4]{Pierre Marchand\orcidlink{0000-0002-4577-8292}}
\author[27,28]{Liton Majumdar\orcidlink{0000-0001-7031-8039}}
\author[29]{Sabyasachi Pal\orcidlink{0000-0003-2325-8509}}
\author[13]{Maria Elisabetta Palumbo\orcidlink{0000-0002-9122-491X}}
\author[16]{Jaime E. Pineda\orcidlink{0000-0002-3972-1978}}
\author[20]{Manoj Puravankara\orcidlink{0000-0002-3530-304X}}
\author[5]{Elena Redaelli\orcidlink{0000-0002-0528-8125}}
\author[3]{Victor M. Rivilla\orcidlink{0000-0002-2887-5859}}
\author[21]{Basmah Riaz\orcidlink{0000-0003-3863-4052}}
\author[22]{\'Alvaro S\'anchez-Monge\orcidlink{0000-0002-3078-9482}}
\author[16]{Silvia Spezzano\orcidlink{0000-0002-6787-5245}}
\author[23]{Leonardo Testi\orcidlink{0000-0003-1859-3070}}
\author[20]{Himanshu Tyagi\orcidlink{0000-0002-9497-8856}}
\author[24,5]{Claudia Toci\orcidlink{0000-0002-6958-4986}}
\author[25]{Alessio Traficante\orcidlink{0000-0003-1665-6402}}
\author[13]{Grazia Umana\orcidlink{0000-0002-6972-8388}}
\author[4]{Charlotte Vastel\orcidlink{0000-0001-8211-6469}}   
\author[26]{Susanne Wampfler\orcidlink{0000-0002-3151-7657}}

\affiliation[1]{INAF, Osservatorio Astrofisico di Arcetri, Largo E. Fermi 5, I-50125, Firenze, Italy}
\emailAdd{eleonora.bianchi@inaf.it}
\affiliation[2]{Leiden Observatory, Leiden University, P.O. Box 9513, 2300 RA Leiden, The Netherlands}
\emailAdd{bouvier@strw.leidenuniv.nl}
\emailAdd{claudio.codella@inaf.it}
\affiliation[3]{Centro de Astrobiología (CAB), CSIC-INTA, Ctra. de Ajalvir Km. 4, 28850, Torrejón de Ardoz, Madrid, Spain}
\emailAdd{lcolzi@cab.inta-csic.es}

\affiliation[4]{Univ. Toulouse, CNES, CNRS, IRAP, Toulouse, France}
\emailAdd{audrey.coutens@utoulouse.fr}

\affiliation[5]{European Southern Observatory, Karl-Schwarzschild-Str 2, 85748, Garching, Germany}
\emailAdd{Marta.DeSimone@eso.org}

\affiliation[6]{Leiden Institute of Chemistry, Gorlaeus Laboratories, Leiden University, PO Box 9502, 2300 RA Leiden, The Netherlands}
\emailAdd{j.enrique.romero@lic.leidenuniv.nl}

\affiliation[7]{Observatorio Astronómico Nacional (OAN), C. Alfonso XII, 3, 28014 Madrid, Spain}
\emailAdd{g.esplugues@oan.es}

\emailAdd{francesco.fontani@inaf.it}

\affiliation[8]{INAF, Istituto di Radioastronomia, Via Gobetti 101, I-40129, Bologna, Italy}
\emailAdd{antonio.garufi@inaf.it}

\emailAdd{lisa.giani@inaf.it}

\affiliation[9]{I. Physikalisches Institut, Universit\"at zu K\"oln, Z\"ulpicher Str. 77, D-50937 K\"oln, Germany}
\affiliation[10]{Max-Planck-Institut für Radioastronomie, Auf dem Hügel 69, 53121, Bonn, Germany}
\emailAdd{ajacob@ph1.uni-koeln.de}

\emailAdd{ijimenez@cab.inta-csic.es}

\emailAdd{marco.padovani@inaf.it}

\emailAdd{linda.podio@inaf.it}

\affiliation[11]{Departament de Química, Universitat Autònoma de Barcelona, 08193 Bellaterra, Catalonia, Spain}
\emailAdd{Albert.Rimola@uab.cat}

\emailAdd{p.riviere@oan.es}

\emailAdd{giovanni.sabatini@inaf.it}

\affiliation[12]{Department of Astrophysics, University of Vienna, Türkenschanzstrasse 17, 1180 Vienna, Austria}
\emailAdd{andrea.socci@univie.ac.at}

\affiliation[13]{INAF – Osservatorio Astrofisico di Catania, via Santa Sofia 78, 95123, Catania, Italy}
\emailAdd{riccardo.urso@inaf.it}

\affiliation[14]{SKA Observatory, Jodrell Bank, Lower Withington, Macclesfield SK11 9FT, UK}
\emailAdd{Tyler.Bourke@skao.int}

\affiliation[15]{Departament de Física Quàntica i Astrofísica (FQA), Institut de Ciències del Cosmos (ICCUB), Universitat de Barcelona, Institut d’Estudis Espacials de Catalunya (IEEC), Spain}
\emailAdd{gbusquet@fqa.ub.edu}

\affiliation[16]{Center for Astrochemical Studies, Max Planck Institute for Extraterrestrial Physics, D-85748 Garching, Germany}
\emailAdd{caselli@mpe.mpg.de}
\emailAdd{jpineda@mpe.mpg.de}

\affiliation[17]{Univ. Grenoble Alpes, CNRS, IPAG, F-38000 Grenoble, France}
\emailAdd{cecilia.ceccarelli@univ-grenoble-alpes.fr}

\affiliation[18]{National Astronomical Observatory of Japan, Japan}
\emailAdd{tomoya.hirota@nao.ac.jp}

\affiliation[19]{School of Physics and Astronomy, University of Leeds, Leeds, UK, LS2 9JT, UK}
\emailAdd{J.D.Ilee@leeds.ac.uk}

\emailAdd{lattanzi@mpe.mpg.de}

\emailAdd{manuela.lippi@inaf.it}

\emailAdd{ana.lopez-sepulcre@univ-grenoble-alpes.fr}

\emailAdd{pierre.marchand.astr@gmail.com}

\emailAdd{maria.palumbo@inaf.it}

\emailAdd{jpineda@mpe.mpg.de}

\affiliation[20]{Tata Institute of Fundamental Research, Homi Bhabha Road, Mumbai 400005, India}

\emailAdd{mpuravankara@gmail.com}

\emailAdd{elena.redaelli@eso.org}

\emailAdd{vrivilla@cab.inta-csic.es}

\affiliation[21]{Universitäts-Sternwarte München, Ludwig Maximilians Universität, Germany}
\emailAdd{basriaz@gmail.com}

\affiliation[22]{Institut de Ci\`encies de l'Espai (ICE), CSIC, Campus UAB, Carrer de Can Magrans s/n, E-08193, Bellaterra (Barcelona), Spain}
\emailAdd{asanchez@ice.csic.es}

\emailAdd{spezzano@mpe.mpg.de}

\affiliation[23]{Dipartimento di Fisica e Astronomia “Augusto Righi”, Università di Bologna, Italy}
\emailAdd{leonardo.testi@unibo.it}

\affiliation[24]{Departamento de Fisica aplicada III, ETSI Universidad de Sevilla,  Camino de los Descubrimientos, 41092 Sevilla, Spain}
\emailAdd{ctoci@us.es}

\affiliation[25]{INAF—Istituto di Astrofisica e Planetologia Spaziali (IAPS), Via Fosso del Cavaliere 100, I-00133 Roma, Italy}
\emailAdd{alessio.traficante@inaf.it}

\emailAdd{grazia.umana@inaf.it}

\emailAdd{charlotte.vastel@utoulouse.fr}

\affiliation[26]{Center for Space and Habitability, University of Bern, Gesellschaftsstrasse 6, 3012 Bern, Switzerland}
\emailAdd{susanne.wampfler@unibe.ch}

\affiliation[27]{National Institute of Science Education and Research, Jatni 752050, Odisha, India}
\affiliation[28]{Homi Bhabha National Institute, Training School Complex, Anushaktinagar, Mumbai 400094, India}
\emailAdd{dr.liton.majumdar@gmail.com}

\affiliation[29]{Midnapore City College, India}
\emailAdd{sabya.pal@gmail.com}

\abstract{About 350 molecules have been identified in the interstellar medium (ISM), including complex molecules relevant to prebiotic chemistry.  A remarkable level of molecular diversity has been observed from the earliest stages of star formation, providing the initial chemical inventory inherited by planetary systems. Radio observations have played a pivotal role in these discoveries, starting with the identification of the first polyatomic molecule, NH$_{\rm 3}$ \citep{Cheung1968}. (Sub-)millimeter observations have revealed complex organic molecules of prebiotic relevance, including formamide (NH$_2$CHO), glycolaldehyde (CH$_2$OHCHO), and even urea ((NH$_2$)$_2$CO), and hydroxylamine (NH$_2$OH), which are possible precursors of RNA nucleotides \citep{Ceccarelli2023, Jimenez2020}. 

However, in dense protostellar regions, dust opacity hampers the detection of molecular emission. Additionally, large molecules and those containing heavy atoms, which have rotational transitions at lower frequencies, often remain inaccessible to current instruments.
The Square Kilometre Array Observatory (SKAO) will provide an unprecedented combination of sensitivity and angular resolution at radio wavelengths. This will allow for the detection of prebiotic species and offer new insights into the chemical pathways that shape emerging planetary systems \citep{Jimenez2022}.

This chapter details the scientific questions and advancements that the SKAO, and more specifically, SKA-Mid equipped with the Band 5 receivers, will pursue in the field of astrochemistry, focusing on the chemical complexity in both high-mass and solar-type star-forming regions.
}

\begin{document}
\include{journal-names}

\maketitle

\section{Introduction: the building blocks of life}
The fundamental question of the origin of life is a key driver of research and lies at the very foundation of astrochemistry. This highly interdisciplinary field investigates how simple atoms in the Universe evolve as building blocks to form prebiotic molecules. Since the discovery of the first polyatomic molecule in interstellar space, it has been clear that the interstellar medium (ISM) is surprisingly rich in molecules, despite its harsh conditions. Interstellar complex organic molecules (iCOMs) are a particularly interesting class of complex molecules \citep[e.g.,][]{Ceccarelli2023, Herbst2009}. These organic molecules, which contain six or more atoms, are significant because of their prebiotic relevance and they have been primary detected in low-mass (M$\sim$1 M$_{\odot}$) and high-mass (M $>$ 8 M$_{\odot}$) star forming regions in our Galaxy
\citep[e.g.,][and references therein]{Jimenez2025}.

\setcounter{footnote}{0}

\begin{figure}[h] 
    \centering
    \includegraphics[width=0.9\columnwidth]{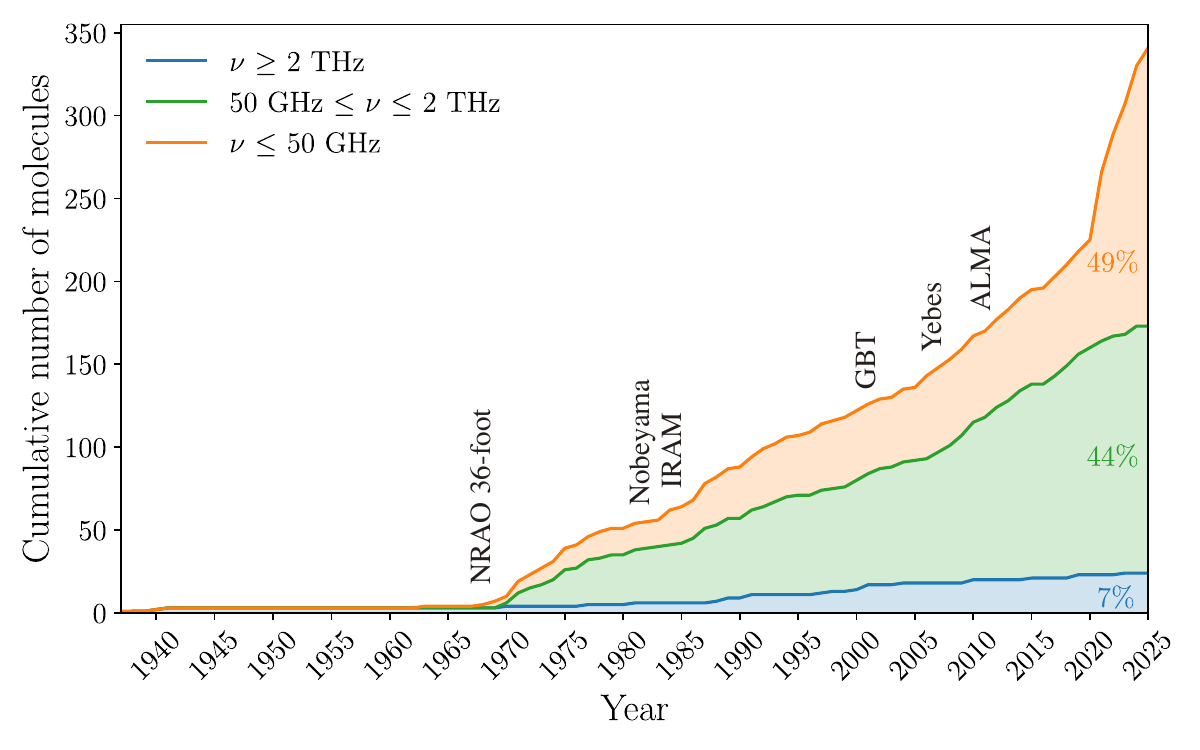}
    \caption[Molecules per year based on the database by Mitsunori Araki]{Molecules per year based on the database\footnotemark\ by Mitsunori Araki, Max-Planck-Institut für extraterrestrische Physik. Figure adapted and updated from \protect\citet{McGuire2022}.}
    \label{fig:molecules-per-year}
\end{figure}
\footnotetext{\url{https://molecules-in.space/}}

A low-mass, solar-type protostar and its planetary system form when a cold ($\lesssim$ 10 K) and dense (> 10$^4$ cm$^{-3}$) core starts to collapse under its gravity \citep[e.g.,][]{Stahler-Palla2004}. During the gravitational collapse, the material accretes asymmetrically from the large-scale envelope ($\sim$ 10$^4$ au) to the central protostellar source forming accretion streamers \citep{Pineda2023}. The conservation of angular momentum leads to the formation of a protostellar disk ($\lesssim$ 100 au) and the ejection of bipolar collimated high-velocity jets and low-velocity outflows. As time progresses, the envelope dissipates, and a planetary system forms from the protoplanetary disk in a process that lasts around 10 Myr \citep[e.g.,][]{Drazkowska2023}. High-mass stars are characterized by spectral types earlier than B3, luminosities exceeding 10$^3$ L$_{\rm \odot}$, and masses greater than 8 M$_{\rm \odot}$.
To explain their formation, several theoretical models have been proposed \citep[e.g.,][and references therein]{Motte2018,Beuther2025}. As examples selected among the most recent ones: (1) the inertial-infall model \citep{Padoan2020}, and (2) the global hierarchical collapse \citep{Vazquez2019}.
The earliest phase involves massive starless clumps or cores, often referred to as infrared-dark clouds (IRDCs). These evolve into high-mass protostellar objects (HMPOs) or hot molecular cores (HMCs). 
The final stage of the process leads to the emergence of classical H\,{II} regions, preceded by the formation of what are known as hypercompact H\,{II} regions.

Thanks to advancements in observational facilities, simple and complex molecules have been identified not only in star- and planet-forming regions and solar system bodies, but also in other galaxies \citep[e.g.][]{Henkel1987, Muller2011,Qiu2018, martin2021,Bouvier2025}.  While high-mass star-forming regions are ideal for new detections due to their high luminosity ($\gtrsim$ 10$^3$ L$_{\rm \odot}$) and gas column densities ($\gtrsim$ 10$^{24}$ cm$^{-2}$), low-mass star-forming regions are in closer proximity ($<$ 0.5 kpc), allowing us to study the spatial distribution and kinematics of molecules to investigate their origin within the protostellar system. Simultaneously, new techniques for laboratory experiments and theoretical computations have been developed to study molecule formation in both ice and gas phases under ISM conditions \citep{Ceccarelli2023}. 
Figure \ref{fig:molecules-per-year} reports the cumulative number of discovered interstellar molecules over time as a function of the year. The recent and dramatic increase in the number of discovered species correspond to the advent of sensitive observational facilities. Figure \ref{fig:molecules-per-year} also shows that radio observations are crucial, as around 50\% of the new detections were made at frequencies below 50 GHz.

Despite the impressive progresses made in recent years, several key questions remain open: What is the maximum level of molecular complexity that can be reached in space? How are these complex species synthesized? To what extent is this complexity inherited by forming planets, and how does this organic material contribute to the emergence of life? Finally, does the chemical diversity observed in the early stages of star formation lead to different planetary compositions?

The Square Kilometre Array Observatory (SKAO) will be pivotal in addressing these open questions. The unprecedented combination of sensitivity and angular resolution at centimeter wavelengths from SKA-Mid, equipped with Band 5 receivers, will allow for the detection of prebiotic molecules and a detailed study of their formation mechanisms during the early stages of star and planet formation. 
The present chapter provides an overview of our current understanding of molecular complexity in star-forming regions, as revealed by sub-millimeter and radio observations, and identifies their inherent limitations (Section \ref{sec:Previous-results}). We highlight the major scientific advances that will be made possible by SKA-Mid in its final AA4 configuration in Band 5 in the detailed study of the molecular complexity of low-mass star-forming regions (Section \ref{sec:low-mass}) and high-mass star-forming regions (Section \ref{sec:high-mass}), providing insights into optimal observational strategies. The chapter also discusses the need to improve analysis tools to handle large datasets and highlights the pressing need for focused laboratory experiments and theoretical computations to provide both spectroscopic data for new species and to study their formation mechanisms (Section \ref{sec:tools&chemistry}).
Finally, the scientific potential for astrochemistry offered by a proposed SKA-Mid Band 6 extension is discussed (Section \ref{sec:future-band6}).

\section{Previous results \& current limitations}\label{sec:Previous-results}
Over the last 15 years, significant work has been done to chemically characterize high- and low-mass star-forming regions through systematic, unbiased spectral surveys and dedicated large programs that cover large frequency ranges. These observations span from infrared to radio wavelengths and have explored both the gas-phase and ice composition.
Notable examples include: QUIJOTE (Q-band Ultrasensitive Inspection Journey to the Obscure TMC-1 Environment; \citealt{Cernicharo2021}), GOTHAM (GBT Observations of TMC-1: Hunting for Aromatic Molecules; \citealt{McGuire2020}),
CHESS (Chemical HErschel Surveys of Star forming regions; \citealt{Ceccarelli2010}), TIMASS (The IRAS 16293-2422 millimeter and submillimeter spectral survey; \citealt{Caux2011}), ASAI (Astrochemical Surveys At IRAM; \citealt{Lefloch2018}), SOLIS (Seeds Of Life In Space; \citealt{Ceccarelli2017}), 
PILS (Protostellar Interferometric Line Survey; \citealt{Jorgensen2016}), CALYPSO (Continuum And Lines in Young ProtoStellar Objects; \citealt{Belloche2020}), FAUST (Fifty AU STudy of the chemistry in the disk/envelope system of Solar-like protostars, \citealt{Codella2021}), ASHES (The ALMA Survey of 70~$\mu$m Dark High-mass Clumps in Early Stages; \citealt{Sanhueza19}), ALMAGAL (The ALMA evolutionary study of high-mass protocluster formation in the Galaxy; \citealt{Molinari2025}), EMOCA (Exploring molecular complexity with ALMA; \citealt{Belloche2016}), GUAPOS (Unbiased ALMA sPectral Observational Survey; \citealt{Mininni2023}), CoCCoA (Complex Chemistry in hot Cores with ALMA; \citealt{Chen2023}), ECHOS (Evolution of Chemistry in the envelope of HOt corinoS; \citealt{Esplugues2023, Esplugues2024}); Ice Age \citep{mcclure23}, JOYS (JWST Observations of Young protoStars; \citealt{vanDishoeck2025}), MINDS (MIRI mid-INfrared Disk Survey; \citealt{Kamp2023}).
These observations have unveiled an astonishing level of molecular complexity, including iCOMs, complex carbon chains and rings, and highly deuterated species. While these studies have enabled a comprehensive census of this molecular complexity and provided insights into the formation mechanisms of some complex species, they have also opened several new key questions in the field.

One surprising result was the discovery of a chemical diversity among protostellar systems. Some sources show emission from iCOMs in a region of $\lesssim$100 au where the temperature is $\geq$ 100 K and are called "hot corino" sources \citep{Ceccarelli2004,Ceccarelli2023}. Others are devoid of iCOMs, with a larger ($\sim 2000$ au) lower temperature (T $\sim$ 30 K) zone within the envelope, enriched in unsaturated (long) carbon chain molecules (e.g., HC$_{\mathrm{2n+1}}$N, C$_{\mathrm{n}}$H,C$_{\mathrm{n}}$H$_2$). These sources were named "Warm Carbon Chain Chemistry" (WCCC) objects \citep{Sakai&Yamamoto2013}. Several protostars were found to possess both a hot corino and a WCCC zone (e.g. L483; \citealt{Hirota2009, Imai2016, Jacobsen2019}), and are thus considered as hybrid objects. This chemical differentiation is likely linked with the chemical composition of the icy mantle coating the dust grains formed at the previous stage within the star formation process, i.e. at the prestellar core stage. While a CH$_3$OH-rich ice mantle at the prestellar stage would favour the formation of a hot corino at the protostellar stage, a C$_4$H-rich mantle would favour the formation of a WCCC object. The observed differences in ice mantle composition may be linked to variations in the duration and physical conditions during the prestellar phase \citep{Sakai&Yamamoto2013, Aikawa2020}. However, the precise origin of this differentiation remains unclear, and it likely exerts an impact on the final chemical composition of the planetary systems forming from these sources. The PErseus Alma CHEmistry Survey \citep[PEACHES,][]{Yang2021} and the ORion Alma New GEneration Survey \citep[ORANGES,][]{Bouvier2022} were designed with same sensitivity and spatial resolution, to compare the chemical nature of solar-mass protostars located in two different star forming regions, characterized by a different environment.
Their main results showed that while hot corinos are abundant in Perseus ($\sim56$\% ; \citealt{Yang2021}), they are much scarcer ($\sim26$\%; \citealt{Bouvier2022}) in Orion. This indicated the possible role of the local environment on the chemical nature of solar-mass protostars. Unfortunately, no WCCC objects were detected in the two surveys, due to spatial filtering affecting the emission of carbon chains, and the scarcity of long carbon chain species falling in the (sub-)mm range, highlighting the need for centimeter studies of carbon chain species.

In addition to the lack of rotational transition of long carbon chain species, two main limitations of (sub-)mm observations have emerged: 1) line confusion and 2) high dust opacity. 
Line confusion arises due to the high sensitivity of modern facilities, which allows for the detection of numerous transitions from both abundant and less-abundant species. This leads to a high probability that observed lines are blended, especially in sources with broad line profiles ($>$ 2-5 km s$^{-1}$), like hot cores and hot corinos. Line confusion significantly limits the ability to identify specific lines, especially those from low-abundance species, and retrieve information from spectra. However, radio observations have a key advantage: they experience less line confusion because fewer transitions from lighter and abundant molecules fall within this spectral range.

\begin{figure}
    \centering
    \includegraphics[width=0.95\linewidth]{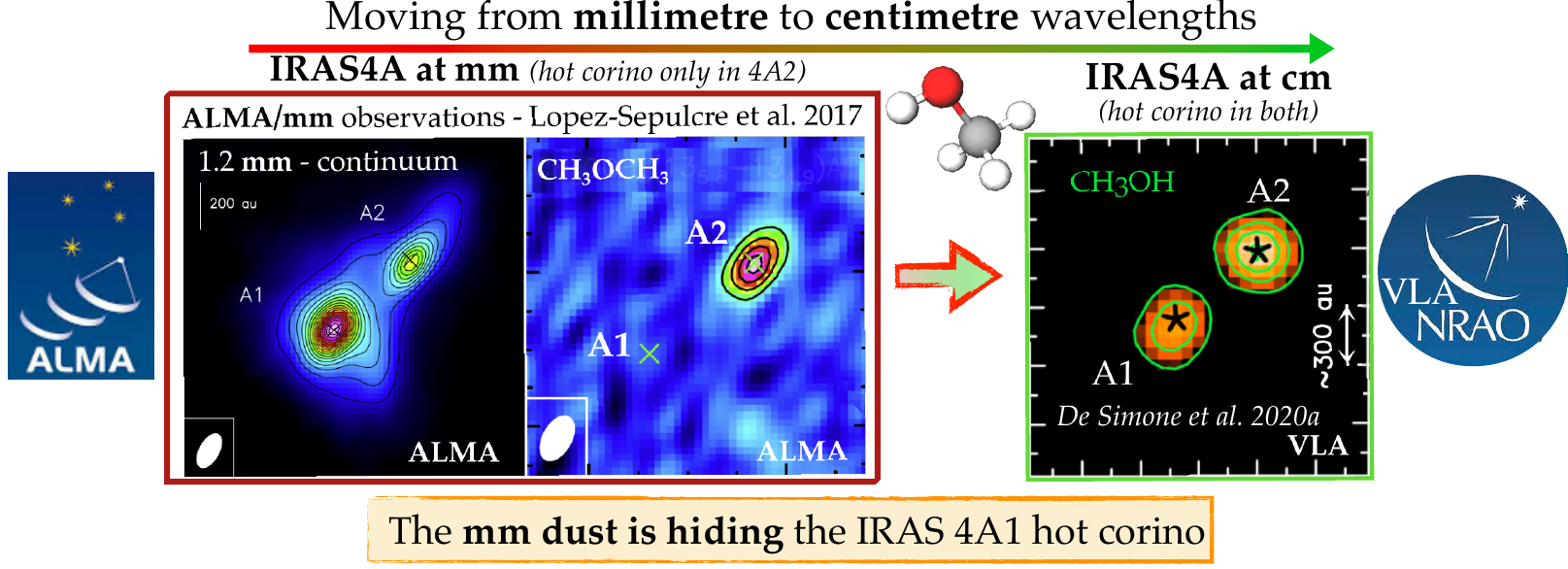}
    \caption{IRAS4A at mm (left, ALMA 1.2 mm continuum and CH$_3$OCH$_3$ emission as representative for all iCOMs) and cm (right, VLA CH$_3$OH emission) wavelengths. The optically thick mm dust obscures the 4A1 hot corino that pops up in the cm and absorbs part of the emission lines in 4A2 \citep{DeSimone_2020ApJL}.}
    \label{fig:dust-opacity}
\end{figure}

Moreover, high dust opacity at millimeter wavelengths strongly affects the ability to observe line emission. Optically thick dust emission can obscure molecular emission, either partially or entirely, which compromises the derivation of physical properties like temperature and column density \citep{DeSimone_2020ApJL, Frediani2025}. Observations of the methanol 25 GHz K-ladder with the Karl G. Jansky Very Large Array (VLA) in the binary system NGC 1333 IRAS 4A revealed methanol emission from both sources \citep{DeSimone_2020ApJL}, which was contrary to what was seen at millimeter wavelengths \citep{LopezSepulcre2017}. They showed that millimeter dust was the culprit, completely obscuring the hot corino in IRAS 4A1 and partially absorbing the molecular emission in IRAS 4A2 (Figure \ref{fig:dust-opacity}). These findings highlight the importance of radio observations for a correct estimation of both molecular column density and gas temperature.
Unfortunately, below approximately 15 GHz, the detection of molecular lines in protostellar environments with the VLA is strongly impeded by sensitivity limits, which restricts observational studies primarily to continuum emission \citep{Coutens2019}.

\section{Low-mass star forming regions}\label{sec:low-mass}

\subsection{SKA-Mid science case for starless and prestellar cores}
Starless and prestellar cores represent the initial stage in the formation of a Sun-like star. Starless cores are cold ($\sim$ 10 K) and dense ($\sim$ 10$^5$ cm$^{-3}$) regions of gas, while prestellar cores are those where signs of gravitational collapse have been observed \citep[e.g.,][]{Tafalla1998, Crapsi2007}. Despite their young age ($<$ 10$^{4}$ yr) both starless and prestellar cores exhibit a rich chemistry, including iCOMs
such as methanol and acetaldehyde \citep[e.g.,][]{Vastel2014, Fuente2019, Jimenez2016,Jimenez2021, Scibelli2020, Rodriguez-Baras2021, Esplugues2022,Esplugues2025, Spezzano2022, Scibelli2024, Cabezas2025, Tasa-Chaveli2025}, as well as complex carbon species \citep[e.g.,][]{Little1977, Suzuki1992, Sakai&Yamamoto2013, Taniguchi2024}. 

\begin{figure}[h]
    \centering
	\includegraphics[width=1\columnwidth]{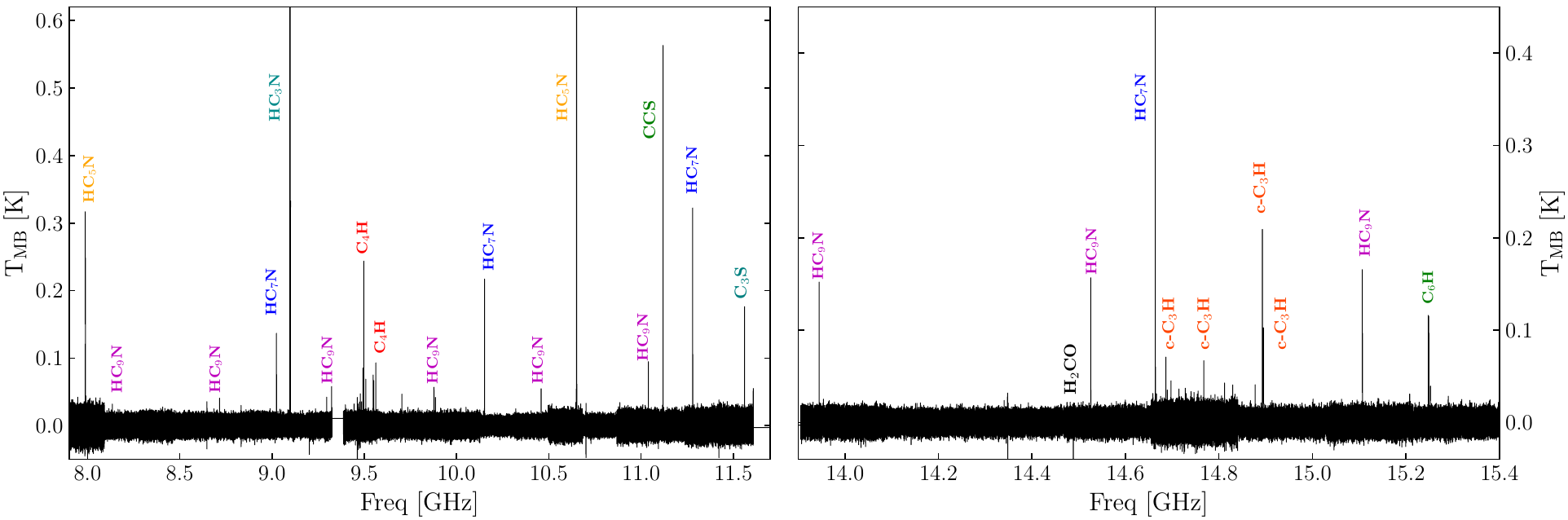}
    \caption{Full spectra (in main-beam temperature, T$_{\rm MB}$) of the L1544 prestellar core observed with GBT between 7.8 and 15.41 GHz, corresponding to the future SKA-Mid Band 5. The labels denote the brightest lines associated with the detected species (adapted from \citealt{giani2025b}).}
    \label{fig:L1544}
\end{figure}

The emergence of new low-frequency receivers \citep[e.g.,][]{Tercero2021} has led to a renewed interest in the chemistry of these objects observed through deep radio surveys conducted with the 100 m Robert C. Byrd Green Bank Telescope (GBT) and the Yebes 40 m telescopes \citep[e.g.,][]{McGuire2020, Cernicharo2022, Burkhardt2021, Bianchi2023}. These programs revealed an unexpectedly high level of complexity in carbon-bearing compounds, including large cyanopolyynes such as HC$_9$N and HC$_{11}$N (see Figure \ref{fig:L1544}; \citealt{Loomis2021,Bianchi2023}), cyclic hydrocarbons such as c-C$_9$H$_8$ \citep{Burkhardt2021} and o-C$_6$H$_4$ \citep{Cernicharo2021}, as well as benzonitrile (c-C$_6$H$_5$CN) \citep{McGuire2018, Burkhardt2021, Agundez2023}. These discoveries indicate that we still lack a comprehensive understanding of the extent to which life-relevant material is formed during the cold phase of prestellar cores and subsequently inherited by the forming planetary systems \citep[e.g.,][]{Caselli2012, Drozdovskaya2014,Altwegg2019,Booth2021}. They also emphasize the critical importance of investigating the chemistry of these early stages of star and planet formation, as well as the impact of different physical factors, such as turbulence \citep{Beitia-Antero2024}, on molecular abundances.

SKA-Mid will allow us to perform a full census of complex carbon bearing species, including cyanopolyynes, long carbon chains and rings in prestellar cores and study their spatial distribution. This will allow (1) to complete the chemical budget at the initial stage of star and planet formation which is inherited afterwards and (2) to study the mechanism of formation of these molecules which are still poorly known (see Section \ref{sec:tools&chemistry}).
Spatial distribution of complex species, such as benzonitrile, were only studied in TMC1 and the emission is observed to follow cyanopolyynes emission being extended ($\gtrsim$ 100$\arcsec$) on the cyanopolyynes peak \citep{Cernicharo2023}.

\begin{figure}[h]
    \centering
	\includegraphics[width=1\columnwidth]{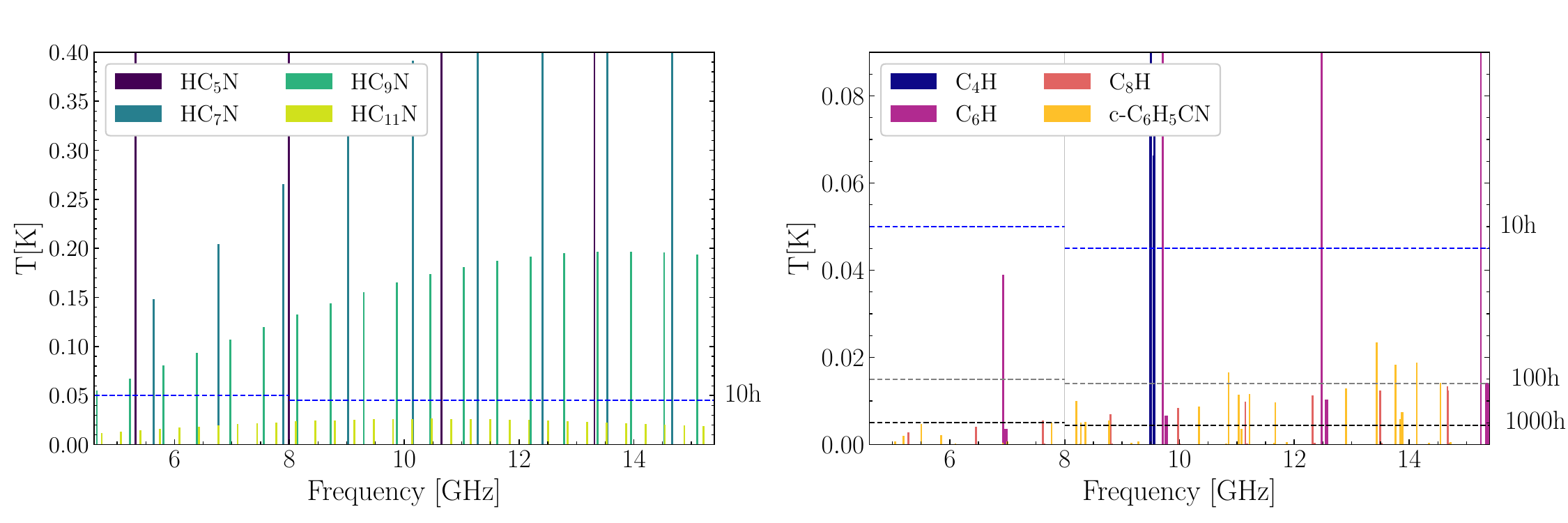}
    \caption{Predicted line intensities of cyanopolyynes (HC$_5$N, HC$_7$N, HC$_9$N, HC$_{11}$N), polyynyl radicals (C$_4$H, C$_6$H, C$_8$H) and benzonitrile (c-C$_6$H$_5$CN) in prestellar cores. The predictions are obtained assuming local thermodynamic equilibrium (LTE), optically thin lines, kinetic temperature (T$_{\rm kin}$) of 8 K, lines full width at half maximum (FWHM) of 0.3 km s$^{-1}$, and extended emission. We assume the following column densities for the different species: N(HC$_5$N) = 4 $\times$ 10$^{13}$ cm$^{-2}$, N(HC$_7$N) = 7 $\times$ 10$^{12}$ cm$^{-2}$, N(HC$_9$N) = 2 $\times$ 10$^{12}$ cm$^{-2}$, N(HC$_{11}$N) = 8 $\times$ 10$^{11}$ cm$^{-2}$, N(C$_4$H) = 3 $\times$ 10$^{13}$ cm$^{-2}$, N(C$_6$H) = 2 $\times$ 10$^{12}$ cm$^{-2}$, N(C$_8$H) = 2 $\times$ 10$^{11}$ cm$^{-2}$, N(c-C$_6$H$_5$CN) = 1.2 $\times$ 10$^{12}$ cm$^{-2}$ (\citealt{Bianchi2023, Cernicharo2021a,giani2025b}). The horizontal dashed lines indicate the 3$\sigma$ sensitivity obtained with SKA-Mid AA4 in the zoom windows. These values were calculated using the SKAO online sensitivity calculator for a spectral resolution of 3.36 kHz ($\sim$ 0.1 km s$^{-1}$) and telescope beams of 45$\arcsec$ $\times$ 43$\arcsec$ and 26$\arcsec$ $\times$ 25$\arcsec$ in Band 5a and Band 5b, respectively. The 3$\sigma$ sensitivity is 50 mK, 15 mK and 5 mK for observing time of 10 hours, 100 hours and 1000 hours, respectively, in Band 5a. It is 45 mK, 14 mK and 4 mK for the same observing times in Band 5b. The values in Band 5b include the sensitivity improvement of a factor $\sim$ 1.4 from the additional 64 MeerKAT antennas which will be equipped with Band 5b receivers.}
    \label{fig:predictions-HCnN-CnH}
\end{figure}

In Figure \ref{fig:predictions-HCnN-CnH} we show predictions for line intensities of polyynyl and cyanopolyynes for typical prestellar cores conditions based on previous single-dish observations.
We assume a kinetic temperature (T$_{\rm kin}$) of 8 K, and column densities for the different species ranging from 2 $\times$ 10$^{11}$ cm$^{-2}$ and 4 $\times$ 10$^{13}$ cm$^{-2}$ \citep{Bianchi2023, Loomis2021, Cernicharo2021a}. We assume extended emission and line widths of 0.3 km s$^{-1}$, based on previous observations. We used the SKAO sensitivity calculator\footnote{https://sensitivity-calculator.skao.int/ Version 2.2.1.} to compute the sensitivity in Band 5. Given the small line widths (FWHM $<$ 1 km s$^{-1}$) typically observed toward prestellar cores, the use of zoom windows is required. We consider a beam size of 45$\arcsec$ $\times$ 43$\arcsec$ and spectral resolution of 3.36 kHz corresponding to 0.14 km s$^{-1}$ to resolve the line profiles. Sensitivity in Band 5b is calculated for a beam size of 26$\arcsec$ $\times$ 25$\arcsec$ and the same spectral resolution corresponding to 0.08 km s$^{-1}$. We also take into account the additional MeerKAT antennas which will be equipped with Band 5b receivers, allowing an improvement in sensitivity of a factor $\sim$ 1.4.
Figure \ref{fig:predictions-HCnN-CnH} (left panel) demonstrates that SKA-Mid AA4 will be able to detect multiple transitions of HC$_{\rm n}$N up to n=11 in several sources sources, resolving the emission both spatially and spectrally, with only 10 hours of observation. With the same amount of time we will also have access to transitions of C$_{\rm 4}$H and C$_{\rm 6}$H, important precursors for cyanopolyynes formation \citep{Giani2025}.
Longer observing time will also enable the discovery of more complex species for the first time, should they be present. For instance, if we assume HC$_{\rm 13}$N is 2.5 times less abundant than HC$_{\rm 11}$N (based on the HC$_{\rm 9}$N/HC$_{\rm 11}$N ratio), a detection would be achievable with 100 hours of observation. Figure \ref{fig:predictions-HCnN-CnH} (right panel) also shows that with 100 hours of integration we would be able to target multiple lines of benzonitrile (c-C$_6$H$_5$CN) and with 1000 hours we will detect C$_{\rm 8}$H. This represents a major step forward for the study of complex carbon bearing species, considering that HC$_{\rm 11}$N has previously been detected in only one source (TMC1) and benzonitrile in a handful of sources through deep integration ($>$500 hours) and stacking techniques.

These observations will, for the first time, unveil the presence of complex carbon chains and rings at the onset of star formation and clarify their underlying formation and destruction mechanisms.

Furthermore, we note that the first AA* configuration will still be capable of detecting multiple transitions of long carbon chains (HC$_{\rm 5}$N, HC$_{\rm 7}$N, HC$_{\rm 9}$N, C$_{\rm 4}$H and C$_{\rm 6}$H) with only 10 hours of observation time, though this will be achieved at a lower angular resolution ($\sim$ 30$\arcsec$) than the final array capability.

\subsection{SKA-Mid science case for protostellar envelopes and accretion streamers}
Chemical complexity reaches its maximum during the protostellar stage. Approximately 10$^4$ yr after the beginning of the gravitational collapse, a central protostellar object forms, surrounded by a protostellar disk that generally extends to less than 100 au. This structure remains deeply embedded within a protostellar envelope, which typically spans from hundreds to thousands of au.
The elevated temperatures arising from accretion and ejection processes during the formation of a protostar lead to the desorption of molecules from icy mantles into the gas phase. These desorbed molecules form the first generation of simple and complex species, which subsequently undergo gas-phase reactions to yield a second generation of more complex species \citep{Ceccarelli2023, Tobin2024}.
 In WCCC sources, such as L1527, a rich variety of carbon-bearing species is detected. Single-dish telescopes have observed cyanopolyynes up to HC$_{\rm 9}$N, alongside other complex carbon bearing species like C$_{\rm 4}$H, C$_{\rm 6}$H, C$_{\rm 4}$H$_{\rm 2}$, l-C$_{\rm 3}$H$_{\rm 2}$, C$_{\rm 6}$H$_{\rm 2}$, CH$_{\rm 3}$CCH, CH$_{\rm 3}$C$_{\rm 4}$H trough single-dish telescopes \citep{Sakai2007, Sakai2008, Araki2017, Law2018}. Their column densities typically vary between 10$^{11}$--10$^{13}$ cm$^{-2}$ and their rotational temperatures are in the range 10-20 K. They are expected to be present in a region of approximately 2000 au around the protostar where the temperature is high enough ($\sim$ 30 K) to have methane (CH$_{\rm 4}$) in the gas-phase. However, the spatial distribution of these long carbon chains remains largely unknown because radio interferometers, such as the VLA, lack the necessary sensitivity and imaging fidelity to map their emission. Simple cyanopolyynes (HC$_3$N and HC$_5$N) and hydrocarbons (c-C$_{\rm 3}$H$_{\rm 2}$) have been successfully observed at millimeter wavelengths using ALMA and NOEMA \citep[e.g.,][]{Fontani+2017, Favre+2018, Calcutt2018, Murillo2022}.
In the prototypical Class 0 protostar IRAS16293-2422, HC$_3$N is observed to be present both in the extended envelope and in the compact hot corino region surronding the A and B protostars \citep{Jaber2017, Calcutt2018, Murillo2022}. ALMA observations from the PILS survey specifically measured the column densities of HC$_3$N in both components N(HC$_3$N)= 4.4 $\times$ 10$^{14}$ cm$^{-2}$ in the A protostar and N(HC$_3$N)= 1.8 $\times$ 10$^{14}$ cm$^{-2}$ in the B protostar, respectively.

\begin{figure}[h]
    \centering
	\includegraphics[width=0.48\columnwidth]{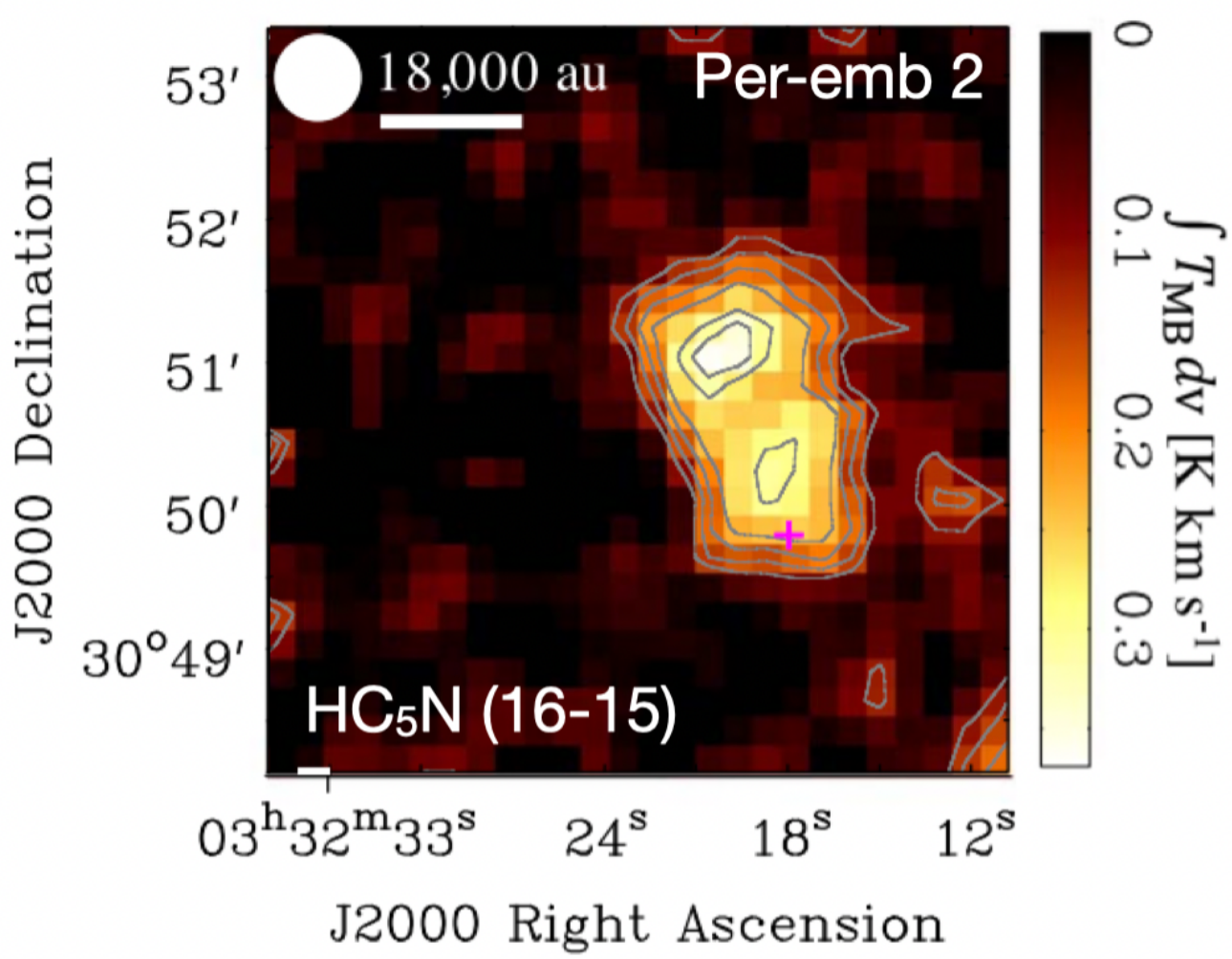}
    \includegraphics[width=0.48\columnwidth]{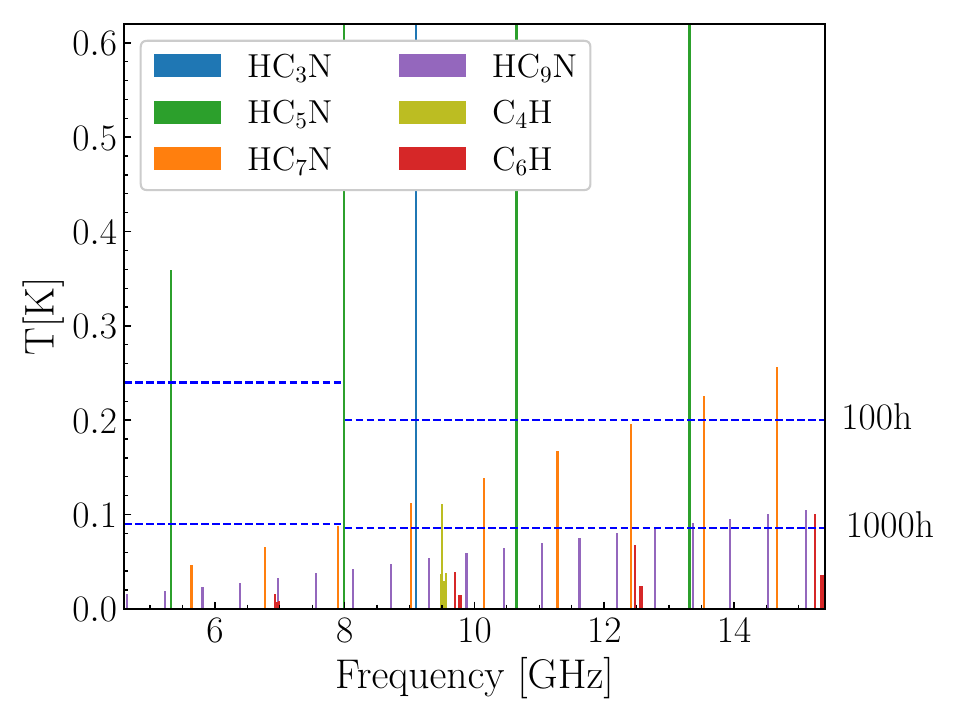}
    \caption{\textit{Left panel:} An accretion streamer detected in HC$_5$N towards the Class 0 young protostar Per-emb-2. Figure adapted from \citet{Taniguchi2024-streamer}. \textit{Right panel:} Predicted line intensities of cyanopolyynes (HC$_3$N, HC$_5$N, HC$_7$N, HC$_9$N), and polyynyl radicals (C$_4$H, C$_6$H) in protostellar envelopes. The predictions are obtained assuming LTE, optically thin lines, kinetic temperature T$_{\rm kin}$=20 K, line FWHMs=1 km s$^{-1}$, and extended emission. We assume the following column densities: N(HC$_5$N) = 1 $\times$ 10$^{14}$ cm$^{-2}$, N(HC$_7$N) = 1.6 $\times$ 10$^{13}$ cm$^{-2}$, N(HC$_9$N) = 4 $\times$ 10$^{12}$ cm$^{-2}$, N(C$_4$H) = 1 $\times$ 10$^{14}$ cm$^{-2}$, N(C$_6$H) = 9 $\times$ 10$^{12}$ cm$^{-2}$ (\citealt{Calcutt2018,giani2025b}). The horizontal dashed lines indicate the 3$\sigma$ sensitivity obtained with SKA-Mid AA4 in the zoom windows. These values were calculated using the SKAO online sensitivity calculator for a spectral resolution of 13.44 kHz ($\sim$ 0.3-0.6 km s$^{-1}$) and telescope beams of 4$\farcs$46 $\times$ 4$\farcs$4 and 2$\farcs$6 $\times$ 2$\farcs$6 in Band 5a and Band 5b, respectively. The 3$\sigma$ sensitivity is 780 mK, 240 mK and 90 mK for observing time of 10 hours, 100 hours and 1000 hours, respectively, in Band 5a. It is 750 mK, 200 mK and 86 mK for the same observing times in Band 5b. The values in Band 5b include the sensitivity improvement of a factor $\sim$ 1.4 from the additional 64 MeerKAT antennas which will be equipped with Band 5b receivers.}
    \label{fig:predictions-protostellar}
\end{figure}

In addition, cyanopolyynes up to HC$_5$N were detected in accretion streamers \citep[][see Figure \ref{fig:predictions-protostellar}]{Murillo2022, Taniguchi2024-streamer}, infalling flows of gas and dust connecting the large-scale envelope to the protostellar disk \citep{Pineda2023}. A common tracer of the streamers so far detected is HC$_3$N \citep{Pineda2020,Tanious2024}, which was used to perform 
the first systematic search for streamers in the star forming region NGC1333 \citep{Valdivia-Mena2024}. The chemical composition of accretion streamers is still unknown but recent studies have shown that it may deeply affect the disk chemical composition, either by delivering non-reprocessed material from the large scale envelope to the disk, or by sputtering the ices in the accretion shocks occurring where the streamer hits the disk \citep{Garufi2022,Podio2024}. 

Beyond their prebiotic interest as crucial ingredients for carbon chain growth in Earth's early chemistry, cyanopolyynes have been used to provide important constraints on the cosmic-ray ionisation rate ($\zeta$), a key parameter for star formation \citep[e.g.][]{Padovani20, Gabici22, Sabatini23, Obolentseva24}.
SOLIS observations revealed a low HC$_3$N/HC$_5$N ratio $(<10)$ in the protostellar cluster OMC-2 FIR4 \citep{Fontani+2017}. Using dedicated astrochemical models, they find that reproducing the observed ratio requires a very high $\zeta$ of the order of $4\times10^{-14}$ s$^{-1}$, in agreement with other studies that used ions and other carbon chains as tracers \citep{Ceccarelli+2014,Favre+2018}. Similarly, \cite{Giani2025} find that in the cold core L1544, HC$_5$N is formed almost entirely by neutral–neutral reactions. However, in warmer gas with higher ionisation fraction, namely the bow shock B1 associated to the protostellar outflow L1157, models underpredict the HC$_5$N abundance unless ion–molecule pathways are taken into account. This implies again a higher $\zeta$, which cannot be explained by invoking the average Galactic cosmic-ray flux. Observations in OMC-2 FIR4 and L1157-B1 have been successfully explained by theoretical models that predict the local acceleration of cosmic rays at the shock fronts of protostellar systems \citep{Padovani+2015,Padovani+2016,Padovani+2021,Lattanzi+2023}. 

SKA-Mid in Band 5 will provide the unique combination of angular resolution and sensitivity in the radio windows to detect complex molecules within protostellar envelopes and investigate their spatial distribution. In Figure \ref{fig:predictions-protostellar} (right panel) we show predictions for line intensities of polyynyl and cyanopolyynes for typical protostellar envelope conditions based on previous single-dish and interferometric observations. We assume extended emission and a kinetic temperature (T$_{\rm kin}$) of 20 K.
 We consider the following column densities N(HC$_5$N)= 1 $\times$ 10$^{14}$ cm$^{-2}$, N(HC$_7$N)= 1.6 $\times$ 10$^{13}$ cm$^{-2}$, N(HC$_9$N)=4 $\times$ 10$^{12}$ cm$^{-2}$.
The column densities are estimated based on HC$_3$N observations from the PILS survey \citep{Calcutt2018}, with other values subsequently rescaled using the cyanopolyyne abundance ratios measured across various sources \citep{giani2025b}. 
 We assume line widths of 1 km s$^{-1}$, based on observations. We used the SKAO sensitivity calculator to compute the sensitivity in Band 5a considering zoom windows, a beam size of 4$\farcs$6 $\times$ 4$\farcs$4 and spectral resolution of 13.44 kHz corresponding to 0.56 km s$^{-1}$. Sensitivity in Band 5b is calculated for a beam size of 2$\farcs$7 $\times$ 2$\farcs$6  and the same spectral resolution corresponding to 0.3 km s$^{-1}$. We also take into account the additional MeerKAT antennas which will be equipped with Band 5b receivers, allowing an improvement in sensitivity of a factor $\sim$ 1.4.
SKA-Mid AA4 will be able to detect one transition of HC$_{\rm 3}$N and multiple transitions of HC$_{\rm 5}$N and HC$_{\rm 7}$N in protostellar sources, with 100 hours of observation. 
The angular resolution of the instrument corresponds to a physical scale of 400–600 au at the distance of the closest star-forming regions. 

This unprecedented resolution will permit us to effectively study the spatial distribution of long carbon chains within protostellar envelopes for the first time. 
Mapping this distribution is crucial, as it provides information not only on their formation mechanisms but also on the survival of organic material within the envelope-to-disk system and its ultimate heritage to planetary system objects like asteroids and comets.

As a comparison, the AA* configuration would require 100 hours of observation time to detect only HC$_{\rm 5}$N at the same angular and spectral resolution. For this science case, the AA4 configuration thus constitutes a major observational advantage.
Longer observation time of 1000 hours with AA4 will also give access to transitions of HC$_{\rm 9}$N, C$_{\rm 4}$H and C$_{\rm 6}$H.
In addition to illuminating the formation routes of long carbon chains and advancing the understanding of WCCC, this study will be instrumental in characterizing the specific imprint left by cosmic rays on cyanopolyyne chemistry, which serves as a powerful diagnostic for the dynamics of star-forming environments.
In fact, by regulating the ionisation fraction, cosmic rays determine the coupling between the gas and the magnetic field.

SKA-Mid observations will probe both the global content of protostellar envelopes and the chemical complexity of shocked gas regions. In these environments, protostellar jets and low-velocity outflows induce sputtering and shattering of dust grains, thereby releasing their molecular content into the gas phase and generating chemically rich regions \citep{Bachiller2001, Codella2010, Codella2017}. The chemistry of protostellar shocks is discussed in the chapter by \citet{Sabatini01.2026.SKA}.
Finally, SKA-Mid, leveraging its full capabilities and unique combination of angular resolution and sensitivity, will be pivotal for exploring the chemical complexity of protostellar disks. A major benefit of observing at radio wavelengths is the ability to probe the disk where dust is less optically thick, providing a direct view into the inner regions critical for planetesimal formation. A dedicated chapter explores the science cases for chemical complexity in planet-forming disks \citep{Podio01.2026.SKA}.


\subsection{A key science case to unveil the chemistry of the early Solar System} 
\begin{figure}[h]
    \centering
	\includegraphics[width=0.9\linewidth]{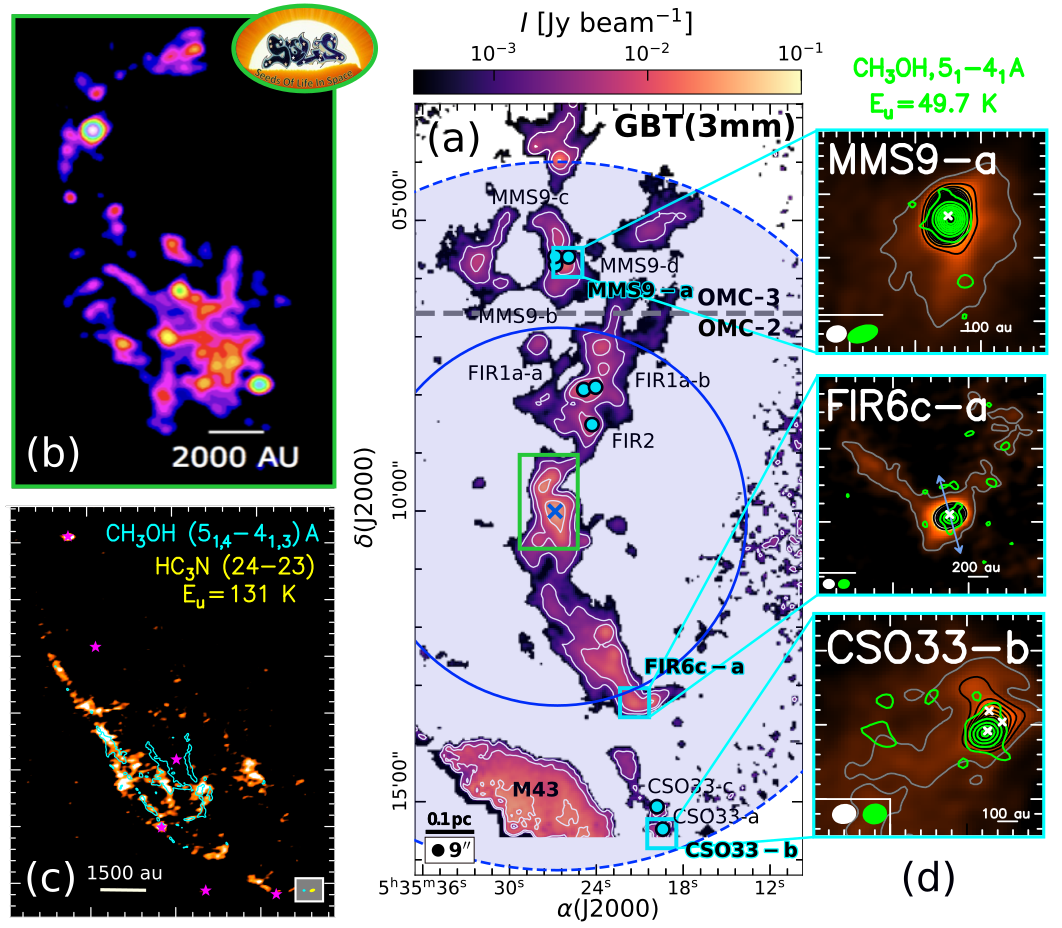}
    \caption{The central panel (Panel a) shows the continuum emission at 3.3 mm of the star-forming filament OMC-2/3, as observed from the MUSTANG camera on the GBT \citep{Mason2020}. The SKA-Mid field of view (FoV) at 8 GHz (dashed blue circle) and 15 GHz (solid blue circle) is shown centered on the FIR4 protocluster, which is highlighted by the green square. Cyan markers indicate the sources observed by the ORANGES survey \citep{Bouvier2021}. Left upper panel (b), reports a zoom of the FIR4 region in continuum emission as observed by the NOEMA SOLIS Large Program \citep{Ceccarelli2017}. 
    The lower-left panel (c) reports a zoomed view of the FIR4 protocluster, observed by ALMA at high angular resolution ($\sim$ 0$\farcs$5) in Band 6 \citep{Chahine2022}. Velocity-integrated emission of the HC$_3$N (24--23) line (E$_{\rm up}$ = 131 K at 218.3 GHz) is shown in colors. Superimposed are contours of the velocity-integrated emission of the CH$_3$OH (5$_{1,4}$--4$_{1,3}$ A) line (E$_{\rm up}$ = 50 K). The methanol contours start at 10 $\sigma$, where $\sigma$ = 12.3 mJy beam$^{-1}$ km s$^{-1}$. The right panel (Panel d) provides a zoom-in on three protostellar sources observed by the ORANGES survey. These observations, taken with ALMA at 1.3 mm, show the continuum emission (in color and gray contours; \citealt{Bouvier2021}), and the CH$_3$OH (5$_{1,4}$--4$_{1,3}$ A) line emission (in green contours; \citealt{Bouvier2022}). SKA-Mid Band 5 observations of a typical star-forming region, such as the OMC, will be able to detect over ten protostellar sources, including already identified hot corinos, in a single setup.
}
    \label{fig:OMC2-Bouvier}
\end{figure}

Crucially, SKA-Mid offers a significantly larger field of view of 6.7$\arcmin$ and 12.5$\arcmin$ at 12.5 GHz (Band 5b) and 6.7 GHz (Band 5a), respectively, compared to the submillimeter and millimeter interferometers (see \href{https://www.skao.int/sites/default/files/documents/Technical%20Information%20Summary%20Sheet%20SKA-AA4.pdf}{SKA-AA4 Technical Summary}). 
As a result, a larger number of sources can be observed with a limited number of pointings. By targeting star-forming regions with high protostellar densities, it is possible to efficiently conduct statistical studies of young stellar objects while maximizing the observing time per pointing, thereby improving sensitivity for each source within the field of view. 
The potential for performing protoplanetary disk demographic studies in nearby regions such as Ophiuchus, Lupus, and Upper Sco is further discussed in the chapter by \citet{Garufi01.2026.SKA}. A use case optimised for astrochemical studies in the Orion star forming regions has recently been proposed\footnote{\href{https://www.skao.int/sites/default/files/documents/d35-SKA-TEL-SKO-0000015-04_Science_UseCases-signed.pdf}{SKA1 Scientific Use Cases}}. 
The Orion Molecular Cloud 2 (OMC-2), at a distance of $393\pm25$ pc \citep{Grossschedl2018-GAIA-Orion}, is an active star-forming filament hosting several star-forming regions with a diverse population of young low- and intermediate-mass protostars and disks \citep{Shimajiri2008, Tobin2019}, including the FIR4 region, which is a well-studied protostellar cluster (see Figure \ref{fig:OMC2-Bouvier}). 
In FIR4, a flux of high-energy cosmic-ray-like particles ionize the molecular gas at a rate exceeding 4000 times the standard value of $1 \times 10^{-17}$ s$^{-1}$ in our Galaxy \citep{Fontani+2017, Favre+2018}, mirroring that experienced by the young Solar System \citep{Gounelle2013}.
In this respect, OMC-2 FIR4 is considered one of the closest analogues to the environment in which our Sun may have formed, making it ideal for studying chemistry reminiscent of our early Solar System. It also has the advantage of being very well studied and to have a wealth of ancillary observation from interferometer such as NOEMA, ALMA and VLA \citep{Ceccarelli2017, Tobin2019}. Interestingly, the ORANGES survey, performed with ALMA, revealed a significant chemical difference compared to other star-forming regions: methanol (CH$_{\rm 3}$OH), typically considered an indicator of hot corino chemistry, was detected in only $\sim$ 26\% of the sources, contrasted with $\sim$ 56\% detected in Perseus \citep{Bouvier2022}. 
This finding has been interpreted as a result of the different local environment: in Orion protostellar feedback, external UV illumination from nearby stars, and local cosmic rays could inhibit the formation of iCOMs, instead favoring the formation of WCCC sources. However, this remains debated because the simple carbon chains detected in the ORANGES survey were insufficient to clearly confirm WCCC sources. In addition, a different hypothesis is proposed to reproduce the carbon-chain abundances measured in TMC-1. This hypothesis relates the observed chemical species to different collapse timescales during the prestellar core stage which would selectively favors the formation of icy dust grain mantles rich in methane (CH$_4$) instead of carbon monoxide (CO; \citealt{Aikawa2020}).

A key science project consisting of deep integration (1000 hours) at moderate angular resolution ($\sim$ 1$\arcsec$--2$\arcsec$) of the OMC2 region will enable transformational science for astrochemical studies and offers significant opportunities for commensal science (see Figure \ref{fig:OMC2-Bouvier}).

Using the SKAO sensitivity calculator, we find that a deep integration of 1000 hours, at 12 GHz, with an angular resolution of 0$\farcs$8 and a spectral resolution of 0.3 km s$^{-1}$, considering the addition of the MeerKAT dishes, will yield a r.m.s. of 18 $\mu$Jy beam$^{-1}$. 

The proposed observations will, for the first time, allow us to study the spatial distribution of long carbon chains and their dependence on the local environment, while overcoming the observational bias introduced by high dust opacity in protostellar sources. This work will finally yield crucial insights into the chemical diversity present in protostellar regions and how this diversity ultimately impacts planet formation, given that planets may inherit a widely varying chemical reservoir. 
The potential for commensality includes, for example, demographic studies of protoplanetary disks and the detection of maser emission across the targeted field. We emphasize that the integration of MeerKAT dishes equipped with the new Band 5b receiver constitutes an important milestone for SKA-Mid to maximise the sensitivity and enable transformational science in astrochemical studies. The potential extension to Band 5a would be strategically invaluable for maximizing the scientific return of SKA-Mid.

\section{SKA science case on astrochemistry of high-mass star formation}\label{sec:high-mass}
\subsection{Deuteration of cyanopolyynes and methanol} 

As in the low-mass regime, the formation of methanol and its deuterated forms (CH$_2$DOH, CH$_3$OD, CHD$_2$OH, etc.) is boosted via surface chemistry processes in the early cold phases, until the energy released by the nascent protostellar object increases the temperature, causing the evaporation of the grain mantles and the release of these molecules into the gas. 
As the gas temperature increases, the deuterated species are expected to get gradually destroyed by the higher efficiency of the backward endothermic reactions \citep[e.g.][]{watson74}. 
Therefore, deuterated fractions of methanol, i.e. the ratio between the column density of the species containing D and that of CH$_3$OH, are expected to be powerful tracers of the evolution of a star forming core also in the high-mass regime.
This theoretical expectation has been confirmed by a few observational works \citep[e.g.][]{fontani2015} that have highlighted a tentative relation between the deuterium fraction of methanol and the evolutionary stage of massive star forming cores, but the statistics is still limited.

Testing these predictions with larger statistics is challenging because the deuterated forms of methanol
are supposed to originate faint lines, which, at (sub-)millimeter wavelengths, can be easily overwhelmed by nearby stronger emission lines of lighter and more abundant molecules. 
Even more challenging is to obtain emission maps of these lines, namely, images at arcsecond resolution.
This observing mode would be particularly important in the high-mass regime, because high-mass star-forming regions are further away from the Sun (heliocentric distances $\geq 1$--$2$~kpc).
For example, at a representative distance of 5 kpc, one would need an angular resolution of 1\arcsec to achieve a linear resolution of 0.024~pc, or 5000 au.
Achieving such resolution would be essential to solve the beam dilution problem (which lowers the chance of detection in single-dish observations) and to properly model the molecular emission, since high-mass star-forming regions usually have a physical and chemical structure that varies on sub-pc (0.01-0.1 pc) linear scales \citep[e.g.][]{gieser21, Redaelli21, Sabatini22, Morii24}.
The combination of high angular resolution and high sensitivity offered by SKA-Mid will
be eminently suitable to solve these problems.
Several transitions of CH$_2$DOH can be observed in the range 1–15 GHz. In this spectral window, the contamination of lines of more abundant molecules is expected to be negligible. 

We can estimate the time required to detect CH$_2$DOH in the band 1--15 GHz by assuming T$_{\rm kin}$=20 K, N(CH$_2$DOH) $= 5 \times 10^{15}$ cm$^{-2}$, source size $=1$\arcsec, and the line FWHM $=1$ km s$^{-1}$. The brghtest CH$_2$DOH lines have an intesity larger than 32 mJy.
The column density assumed is a mean source-averaged value measured towards high-mass protostellar cores \citep{fontani2015}.
At a rest frequency of
5 GHz, assuming the AA$^{*}$ configuration, a velocity resolution of 1 km
s$^{-1}$, the $1 \sigma$ rms noise in the spectrum after 50 hours of integration on source is 0.077 mJy for a S/N ratio larger than
$3\sigma$.

\subsection{Chemical complexity in low-metallicity/galactic environments}

\begin{figure}[h]
    \centering
    \includegraphics[width=1.0\columnwidth]{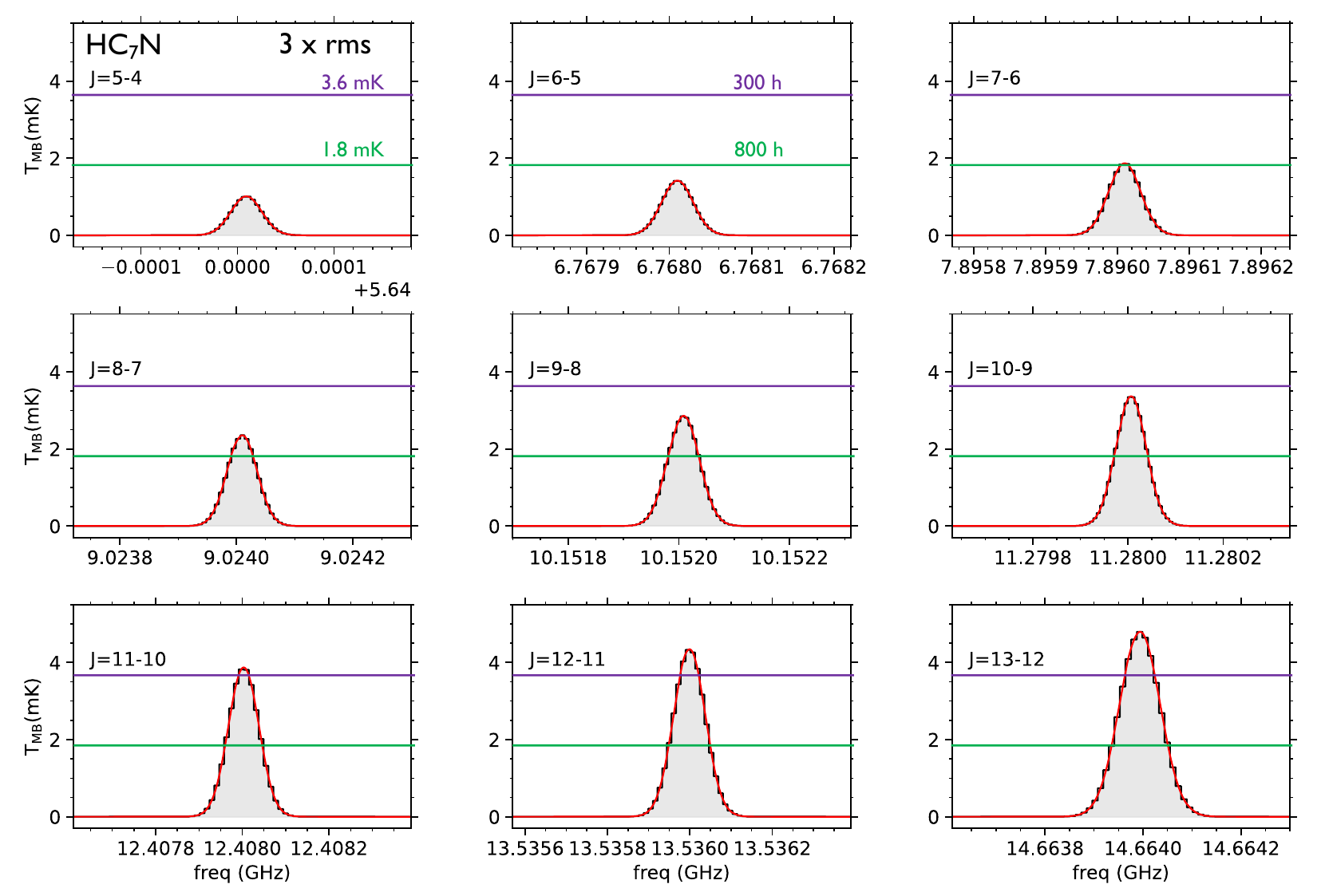}
    \caption{LTE simulated spectra for HC$_{7}$N rotational transitions that will be covered with SKA Mid 5b. The parameters used are $N$= 3$\times$10$^{11}$ cm$^{-2}$ $T_{\rm ex}$ = 11 K and FWHM = 2 km s$^{-1}$. The green and purple horizontal lines represent the 3 rms limit for a total observing time of 800 h and 300 h, respectively. See the text for more information. }
    \label{fig-og}
\end{figure}

The Outer Galaxy (OG), i.e. the portion of the Galactic disk beyond the Solar Circle (i.e. with Galactocentric distance, $R$$_{\mathrm{GC}}$, 9$<$$R$$_{\mathrm{GC}}$$<$24 kpc), is characterised by chemical properties different from those of the Inner Galaxy. 
One of the major differences it that metallicity is lower than the Solar one (up to a factor of four lower at $R$$_{\mathrm{GC}}$ = 19 kpc; \citealt{Fernandez-Martin2017, Wenger2019, Shimonishi2021}).
But other differences crucial for chemistry are expected.
In particular, at distances $>$16 kpc, the far OG is also characterized by lower gas and stellar densities, weaker interstellar radiation fields, and fewer supernova remnants to trigger star formation \citep[e.g.][]{Nakanishi2016, Baharin2025}.
There is also a negligible perturbation from spiral arms.
In the OG, with fewer metals and dust grains, and less
shielding from UV radiation that can destroy fragile molecules, the expected scenario would be that the
formation and survival of molecules is less efficient than in regions close to the Sun. 
However, this scenario is being questioned by growing observational evidence. 
First, the molecular abundances in OG star forming regions do not simply scale with metallicity \citep{Bernal2021,Fontani2022b}; second, iCOMs
are more abundant than
expected in low-metallicity galaxies such as the Large and Small Magellanic Clouds, external dwarf galaxies where ALMA observations have suggested the presence of hot molecular cores \citep{Shimonishi2016, Sewilo2018, HamedaniGolshan2024}.
In particular, these observations have revealed emission of methanol, methyl formate and dimethyl ether, associated with metallicity-scaled abundances compared to those found in local hot cores. 
These iCOMs are thought to be precursors of more complex molecules \citep[e.g.][]{Bohama2014, Zhu2020}, some of them being precursors to biological molecules. 
Altogether, these results suggest the existence of a chemical complexity that might be comparable to that found in the inner region of the Galaxy in spite of the different metallicity. 

The current on-going CHEMOUT project (e.g. \citealt{Fontani2022,Fontani2022b}) aims at studying the CHEMical complexity in star-forming regions of the OUTer Galaxy, considering a sample of 35 dense cores located at 12 $<$$R$$_{\mathrm{GC}}$$<$ 24 kpc. 
Some key complex molecules, such as CH$_{3}$OH and CH$_3$CCH, have already been observed. 
Nevertheless, detection of more iCOMs in these regions is expected based on recent results towards the star-forming region WB89-789 in the outer Galaxy observed with ALMA \citep[]{Shimonishi2021}. 

Heavy complex molecules, especially carbon-bearing compounds such as cyanopolyynes, present a large number of transitions at low ($<$20 GHz) frequencies. In particular, SKA-Mid in Band 5b, covering from 8.3 GHz to 15.4 GHz, will allow the detection of a large number of these type of molecules, widening our understanding of how complex can the chemistry become in such, in principle, harsher conditions in the Outer Galaxy. In addition, these detections will contribute to significantly improve the astrochemical models used to study chemistry in low metallicity environments by constraining the initial elemental abundances and their relative ratios in this region of the Galaxy, providing accurate information about the metallicity levels in the Outer Galaxy not only as a function of radial distance but also azimuthally.

We have estimated the integration time needed to observe cyanopolyynes in the outer Galaxy simulating a spectra starting from the IRAM 30m observations by \citet{Fontani2022} towards the CHEMOUT sample. We have assumed LTE conditions with typical line width (FWHM = 2 km s$^{-1}$), excitation temperature (T$_{\rm ex}$ = 11 K), and column densities (N = 10$^{12}$ cm$^{-2}$, 5 $\times$ 10$^{11}$ cm$^{-2}$, and 3 $\times$ 10$^{11}$ cm$^{-2}$), for HC$_{3}$N, HC$_{5}$N, and HC$_{7}$N, respectively. Moreover, we have smoothed the spectra to 7.2 kHz ($\sim$ 0.3 km s$^{-1}$ at 7 GHz). Using the SKAO online sensitivity calculator in the AA* configuration in Band 5b, with 300 hours of observations we will obtain a spectral surface-brightness sensitivity of 0.02 K with a synthesised beam-size of 8.17$^{\prime \prime}$ at the central frequency of 11.85 GHz. Since the emission of these star-forming regions is diluted in the IRAM 30m beam at 90 GHz (28$^{\prime \prime}$), being the source size of $\sim$ 7$^{\prime \prime}$ or smaller (e.g. \citealt{fontani2024}), we have applied a beam dilution factor of (7$^2$+28$^2$)/7$^2$ =17. This is equivalent to obtain a sensitivity of $\sim$ 1.2 mK in the non-diluted spectra for 300 hours of observations. Doing the same estimate for 800 hours of observations we will obtain $\sim$ 0.6 mK of sensitivity. To summarize, with 300 hours of observations we will be able to detect the rotational transition $J$=1-0 of HC$_{3}$N, as well as HC$_{5}$N (5--4) and (4--3), and HC$_{7}$N (13--12), (12--11), and (11--10) in the OG. To detect lower-intensity transitions $\ge$800 hours of observing time are needed (see e.g. Fig.~\ref{fig-og}). 

Another important improvement is the Field of View (FoV). In fact, with SKA-Mid at 3.3 cm the FoV will be 450\arcsec (33 pc at 15 kpc, compared to 50\arcsec  or 3.6 pc with ALMA at 3 mm). 
Considering that the Orion molecular cloud is 90 pc elongated, with SKA-Mid we will be able to map a cloud similar to that size with only two separated pointings in the OG. Thus, even if the integration time for low-metallicity sources is high, with one or two pointings we will obtain the chemical and physical properties of entire large filaments of star formation in the outer Milky Way.

Regarding the continuum, cm wavelength observations with SKA-Mid will also allow to detect embedded ultracompact HII regions in low metallicity massive star-forming regions. Considering typical massive star-forming regions in the inner Galaxy, \citet{sanchez-monge2011, SanchezMonge2013} detected embedded HII regions with the VLA and ATCA, with a peak of 0.6~mJy in a beam of 9$^{\prime \prime}$ at 3.6~cm. Considering that the OG sources can be up to 5 kpc further away in heliocentric distance with respect to inner Galaxy sources, we will expect maximum one order of magnitude of intensity lower. Thus, to reach a sensitivity of 0.01 mJy/beam at 9 GHz (1 GHz of bandwidth) with a beam of 5$^{\prime \prime}$ we will dected HII continuum sources with 7.4 min in AA* configuration (48 seconds in AA4). While the VLA can now achieve comparable sensitivity in only 11 minutes, the unrivaled imaging fidelity provided by SKA-Mid delivers a qualitative step forward in capability.

SKA-Mid will be the first interferometer that will operate in the cm wavelengths in the southern hemisphere. This will allow to map new sources in the outer Galaxy with declination from -90\textdegree to -45\textdegree, not covered at all with VLA from the northern hemisphere. The new sources can be selected based on their infrared emission from the Hi-Gal sample (\citealt{molinari2010,mege2021,elia2021}), for which some of them in the Galactic quadrants with galactic longitude from 180\textdegree \> to 280\textdegree \> have been also recently observed in CO (2-1) with APEX (\citealt{urquhart2024,urquhart2025}).

Regarding the influence of low metallicity on planet formation, protoplanetary disks in low-metallicity environments exhibit significant physical and chemical differences compared to those with solar metallicity. Observations of young stellar objects in the outer Galaxy \citep[$\rm O/H\sim 1$,][]{Rudolph2006} reveal that disks there have shorter lifetimes, likely due to enhanced mass accretion rates and more effective photoevaporation \citep{Yasui2010,Yasui2016}. Chemical models show that the reduced abundance of metals and dust allows ultraviolet radiation to penetrate deeper into the disk, altering the chemistry by enhancing gas-phase formation efficiencies and shifting molecular snowlines outward \citep{Guadarrama2022}. From a planet formation perspective, lower metallicity limits the amount of solids available for dust growth and pebble formation, reducing the efficiency of the streaming instability mechanism and restricting planetesimal formation to disks with metallicities above [Fe/H] $\rm \sim$ --0.6 \citep{Andama2024}. Together, these findings explain the observed planet–metallicity correlation, in which the frequency  of giant planets increase with stellar metallicity \citep{Gonzalez1997, Fischer2005, Santos2003}. Future observations with SKA-Mid will enable highly sensitive detections of cold gas in faint, metal-poor disks across the Galaxy and nearby dwarf galaxies. Its exceptional spatial resolution and sensitivity will allow detailed mapping of disk chemistry, kinematics, and evolution under low-metallicity conditions, providing critical constraints on how the initial conditions for planet formation vary with metallicity and bridging the gap between chemical models and observed exoplanet demographics.

\section{Advancing astrochemistry beyond observations}\label{sec:tools&chemistry}

Astrochemistry is a highly interdisciplinary field by its nature, which requires strong synergies between astronomical observations and chemistry. We provide below some perspectives on how the new SKA-Mid observations will advance and catalyze related fields within astrochemistry.
\subsection{Laboratory experiments}
Laboratory rotational spectroscopy provides the essential foundation for detecting and identifying complex species in the interstellar medium. As new radio facilities such as the SKA-Mid push molecular astronomy into unprecedented regimes of sensitivity and frequency coverage, precise laboratory measurements become even more critical. 
Laboratory rotational spectroscopy provides the high-precision rest frequencies, line strengths, hyperfine structures, and partition functions needed to securely identify weak, blended, or previously unknown transitions of complex molecules \citep{McCarthy2021, McGuire2022}. Recent detections in the cm–mm domain illustrate the power of this synergy. Molecules such as cyanomethanimine (HNCHCN) and propenal (CH$_2$CHCHO) have been identified in cold sources like G+0.693$-$0.027 and TMC-1 \citep{Zeng2018, McGuire2020}, while methyl isocyanate (CH$_3$NCO), glycolonitrile (HOCH$_2$CN), and amino acetonitrile (NH$_2$CH$_2$CN) have been observed in both hot cores and cold dark clouds \citep{Belloche2019, Zeng2021}. Many of these detections were enabled, or even made possible, by recent advances in laboratory spectroscopy that provided accurate measurements of low-energy rotational transitions \citep{Bizzocchi2017, Puzzarini2017, Endres2016, Spezzano2022}.
Despite this progress, significant gaps remain. Many prebiotically relevant molecules, including amides, nitriles, esters, and unsaturated carbon chains, are still poorly characterized in the cm-wave regime, particularly in their vibrationally excited states, isotopologues, conformers, and larger molecular forms ($>$ 8--10 atoms; \citealt{Endres2016, Puzzarini2020}). Discrepancies between transitions detected at radio wavelengths with the GBT and those reported in spectroscopic databases have been observed even for relatively simpler species, such as c-C$_3$H and C$_6$H \citep{giani2025b}, triggering new spectroscopic measurements \citep{Remijan2023, Xue2025}.

Future laboratory efforts must therefore prioritize:
\begin{itemize}
    \item Extending frequency coverage into the 1--20~GHz range with sub-kHz precision \citep{McCarthy2021}.
    \item Characterizing isotopologues and vibrationally excited states that trace formation pathways and isotopic fractionation \citep{Ceccarelli2023,Puzzarini2017}.
    \item Measuring dipole moments and line strengths for weak or forbidden transitions that become accessible with next-generation sensitivity.
\end{itemize}

Laboratory experiments are essential not only for the spectroscopic characterization of molecular species but also for elucidating the chemical processes involved in the formation of iCOMs.
In particular, experiments allow us to study the formation of molecular species resulting from the exposure of icy grain mantles to ionizing radiation, such as UV photons \citep{allamandola88, munozcaro03} and low-energy (keV–MeV) cosmic rays \citep{kanuchova16}, as well as the interaction between ices and gaseous hydrogen \citep{dulieu19, fedoseev25}.
Available experimental setups enable the formation of icy grain mantle simulants through the deposition of abundant species detected in the solid phase toward star-forming regions, such as H$_2$O, CH$_3$OH, NH$_3$, CO, and CO$_2$, under conditions of very low temperature (T $\sim$ 10 K) and ultra-high vacuum (P $\sim$ 10$^{-10}$ mbar). Ices are then exposed to UV sources (typically hydrogen Lyman-alpha photons at 121 nm), electron beams (a few keV), and ion beams (hundreds of keV to a few MeV). The interaction between molecules in the ice and energetic photons or particles causes changes in the physical and chemical properties of the ice. Reactions are triggered by the transfer of energy and momentum from the ion to the molecules within the ion track, a confined region along the ion path where radicals and molecular fragments are formed \citep{rothard17}. Within timescales of 10$^{-12}$ to 10$^{-9}$ seconds, these excited species recombine to form new compounds \citep{urso22, faure25}. 
When a carbon-bearing molecule is present in the processed ice mixture, its abundance decreases exponentially with increasing dose, that is, the energy deposited in the sample by incident particles. The destruction of precursor species is followed by the formation of new compounds, including iCOMs \citep{bennett05, oberg09, palumbo99, urso17}, carbon chains \citep{palumbo08, urso19}, and even complex carbonaceous structures including aromatics \citep{ferini04}. The abundances of these species, considered the building blocks of the complex matter found in meteorites and other extraterrestrial samples, primarily depend on the ice composition and irradiation dose \citep{urso22}.
In the laboratory, carbon chains and more complex carbonaceous structures can be identified using various analytical tools, including spectrometers (e.g., FT-IR, UV-Vis, Raman) and mass spectrometers (e.g., QMS). In star-forming regions, icy grain mantles can only be directly observed at infrared wavelengths \citep[e.g., ][]{mcclure23}, which, however, pose significant limitations in detecting complex compounds. These species are typically formed in relatively low abundances compared to the main ice components. 
Furthermore, infrared spectra show vibrational transitions of chemical bonds, providing information about bond types and functional groups present in the molecules trapped within the ices. Carbon chains are characterized by repetitive bonding patterns and a limited variety of functional groups, leading to overlapping IR absorption bands. This makes it difficult to distinguish between molecules that share similar bonds and functional groups. 
Thus, although the JWST is providing primary information on the properties of pristine ices, the detection of carbon chains will be hampered by observational constraints.  
However, the release of species formed in ices into the gas phase enables their detection at the wavelengths observable with SKA-Mid. The detection of carbon chains would be highly significant, as it would strengthen the evidence for a link between the chemistry active in the early stages of star formation and the composition of complex organics found in primordial extraterrestrial samples.

Finally, laboratory experiments can elucidate the gas-phase formation of iCOMs. However, fully reproducing the necessary low temperatures and low pressures within a single experimental setup remains a significant technical challenge. Collision-free techniques, such as crossed molecular beam (CMB) experiments \citep{Casavecchia2009}, reproduce the low number densities typical of the ISM and allow the identification of products, the determination of branching fractions, and the characterization of scattering dynamics. In the CMB apparatus, two supersonic beams intersect in an ultra-high vacuum chamber (down to 10$^{-7}$ mbar), ensuring single-collision conditions. The products are analyzed by a rotatable quadrupole mass spectrometer, which measures both their angular and time-of-flight distributions.
The CRESU (Cinétique de Réaction en Écoulement Supersonique Uniforme) technique, instead, reproduces very low temperatures (down to 10 K) and provides absolute rate constants \citep{smith2000}. It uses Laval nozzles to generate a cold supersonic flow from a pressurized gas, and the rate constants are derived from the decay of reactants or the formation of products. However, CRESU does not give direct information on the identity of the products. Only a few setups, such as the Bordeaux apparatus \citep{costes2010} and CRESUSOL \citep{Guillaume2024}, can simultaneously reproduce both temperature and density conditions.
A further difficulty arises because many astrochemical reactions involve radical–radical collisions. In CRESU, rate constants are normally measured under \textit{pseudo-first-order conditions}, where one reactant is in large excess. For radical–radical systems this is not feasible, since neither radical can be produced in sufficient excess, making absolute rate constants inaccessible with current techniques. In these cases, theoretical calculations remain a crucial complement to experiments for the characterization of gas-phase reactions and for improving astrochemical models. 

\subsection{Computational chemistry}
Atomistic modelling, often based on quantum mechanics (QM), is a crucial alternative approach to astrochemical problems. Its relevance lies in its capacity to provide unprecedented, quantitative atomistic-scale information (e.g., molecular structures, chemical energetics, dynamical effects...), which can be uniquely obtained with QM simulations. The key point is that the ISM conditions at the beginning of star formation are characterized by very low temperatures and densities, which makes the chemical processes to be truly individual. This is precisely what QM computations simulate, namely, molecular-level isolated systems. In addition, the data obtained from QM modelling can be fed into astrochemical models (e.g., binding energies, rate constants), thus improving the predictions and hence the confrontability with observations.

Atomistic simulations depend on the state of matter through which the chemical processes take place: in the gas phase or on the surfaces of grains. In the gas phase, QM modelling essentially addresses chemical reactions. However, gas-phase reactivity between two species usually gives rise a wealth of different reactive channels, the potential energy surface (PES) of which need to be characterised. PES provides a molecular-level description of reaction mechanisms and their characteristics, such as energy barriers, reaction energies and vibrational frequencies. Moreover, using PES data, kinetic theories can be applied deriving rate constants and branching ratios, thus determining the viability of the channels under the ISM conditions \citep[e.g.,][]{Skouteris2018,Vazart2020}.

On interstellar grain surfaces, the systems are structurally more complex since the grains themselves need to be considered. In that respect, two different approaches can be adopted \citep{Rimola2021}: (i) periodic models, in which a unit cell, representing the minimal surface building block, is extended in 2D, thus generating an "infinite" surface with a finite thickness; and (ii) cluster models, which mimics the surface as a finite ("molecular") fragment. On these atomistic models, elementary surface phenomena are then simulated, which can be: (i) adsorption/desorption (molecules landing or taking off the grain surface); (ii) diffusion, in which adsorbed species move over the surfaces; and (iii) chemical reaction, in which reactive components encounter each other to react. Chemical reactions are studied in analogy to the gas phase (i.e., PES characterization), but in this case, reactant/surface interactions significantly constraint the process and other possible competitive channels \citep[e.g.,][]{Rimola2014-COhydrogenation,Rimola2018,Perrero2022-ethanol,Enrique-Romero2022,Perrero2023,Perrero2024}. Regarding adsorption/desorption, a crucial parameter is the binding energy (BE), which reflects the strength of the species-to-surface interaction and is directly associated with the ice-to-gas transition during adsorption/desorption events. Thus, BEs determine whether a species remains in the gas or solid phase, influencing the chemical composition in astrophysical environments. In astrochemical models, diffusion also depends on the BE, as diffusion barriers (E$_{\rm dif}$) are commonly assumed to be a fraction of the BE (E$_{\rm dif}$ = \textit{f} $\cdot$ BE). Therefore, BEs are essential input parameters in astrochemical models, in such a way that introducing accurate BEs is crucial for the reliability of these models. Efforts to obtain BEs computationally have evolved over time, from simple ice models of few molecules  \citep[e.g.,][]{Wakelam2017BE,Das2018}, to medium-sized ones (20-33 water molecules; e.g., \citealt{Shimonishi2018-BEatoms,Bovolenta2022-BEdatabase}), into larger, more realistic, extended models \citep[e.g.,][]{Ferrero2020,Perrero2022-SBE,Martinez-Bachs2024}. This has allowed us to move from calculating single values to deriving binding energy distributions, where the most recent and reliable modelling strategy is provided by the ACO-FROST \citep{Germain2022} and PolyCleaver \citep{Mates-Torres2024} procedures.
Both include algorithms to systematically sample surface binding sites and calculate the corresponding BEs \citep{Tinacci2022-ammonia,Tinacci2023-waterBE,Bariosco2024,Bariosco2025,Mates-Torres2025}, resulting in BE histograms that reflect the distribution of all calculated BEs at each site.

In general, atomistic modelling is a matter of balancing accuracy and computational cost. The computational demands of a simulation depend on two main factors: (i) the already mentioned system’s size, that is, the number of nuclei and electrons involved, and (ii) the level of approximation used to solve the quantum mechanical problem (e.g., how is the Schr{\"o}dinger equation solved). Based on these approximations, one can distinguish different families of methods. In astrochemistry, density functional theory (DFT) methods are commonly used due to their favourable compromise between accuracy and efficiency. Rather than explicitly solving for the whole system's wave function, DFT reformulates the problem in terms of the electron density, a simpler quantity that still captures the essential physics of the system. Remarkably, due to the inherent intricacies of density functional approximations, benchmark studies comparing highly accurate methods to the more accessible DFT-based ones are essential to assess the reliability of the results. More accurate approaches may be found in the so-called "wave function-based" methods family, which directly work with approximations to the electronic wave function. In this category, methods like coupled cluster (e.g., CCSD(T)), or the so-called multi-reference ones (e.g., CASPT2, NEVPT2, MRCI...) can reach very high accuracies, but their computational cost increases steeply with system size.
Atomistic modelling is a rapidly evolving field, and many advanced and specialised methods exist beyond those introduced here.

\subsection{Astrochemical models \& chemical networks}

Astrochemical models are essential tools to interpret observations, as they compute the time evolution of molecular abundances starting from given initial conditions and physical parameters. A key component of these models is the reaction network, which accounts for all possible formation and destruction pathways and their associated rate constants. Model predictions depend critically on the completeness and accuracy of these networks: missing or outdated reactions can lead to unreliable results. Quantitative information on chemical processes is therefore indispensable, and both laboratory experiments and theoretical calculations play a central role in providing it. At present, the most widely used astrochemical databases are KIDA \citep[Kinetic Database for Astrochemistry;][]{wakelam2024} and UDfA \citep[UMIST Database for Astrochemistry;][]{Umist2022}. These databases are frequently employed in several astrochemical codes, such as the Willacy Model \citep{Willacy98}; \textsc{Alchemic} \citep{Semenov2010}; \textsc{Magickal} \citep{garrod2011formation}; \textsc{Grainoble} \citep{taquet2012multilayer}; \textsc{Monaco} \citep{Vasyunin2013}; \textsc{Astrochem}\footnote{\url{https://github.com/smaret/astrochem}} \citep{maret2013chemical}; \textsc{krome}\footnote{\url{https://www.kromepackage.org/}} \citep{Grassi14}; the Rokko code \citep{Furuya2015}; \textsc{Nautilus}\footnote{\url{https://astrochem-tools.org/codes/}} \citep{Ruaud2016}; \textsc{UCLCHEM}\footnote{\url{https://uclchem.github.io/}} \citep{holdship2017uclchem}; Sipil{\"a}'s model \citep{sipila2022}; and \textsc{Pegasis} \citep{pegasis}, among others. Despite continuous updates informed by studies of both gas-phase and surface processes \citep[e.g.,][]{balucani2015formation,Rimola2018,Skouteris2018,Vazart2020,Enrique-Romero2022,giani2023revised,Perrero2024}, several limitations remain. Networks often lack reactions involving large molecules, or conversely include processes that should not be present \citep{Tinacci2023gretobape}. Remarkably, nearly 80\% of the gas-phase reactions currently listed have never been investigated experimentally or theoretically. Different studies have focused on developing methods to identify the key reactions that primarily impact astrochemical model predictions \citep{Fernandez-Ruz2023, Fernandez-Ruz2025, Holdship2018, Heyl2022}.
Recent results have shown that the chemistry of carbon chains and cyanopolyynes remains incomplete, and their abundances are not always reproduced across different sources (\citealt{loison2014,Loomis2021,giani2025b}), highlighting the need for coordinated efforts between experimentalists and theoreticians to improve networks reliability.

Given the unprecedented sensitivity and data quality that SKA-Mid observations will afford, the forthcoming results will mandate a thorough and immediate revision of the formation and destruction pathways for these molecular species. 

This preparatory work is essential for accurately interpreting the new data and maximizing the scientific return of the instrument.

\section{Conclusions and future perspectives on SKA-Mid band 6}\label{sec:future-band6}

The SKAO will inaugurate a new era of astrochemical studies at high angular resolution in the radio domain. It will enable, for the first time, the detection of new complex carbon-bearing species relevant to the origin of life. Simultaneously, it will grant access to physical regions around young protostars, the planet-forming regions, that are currently hidden by the high dust opacity. 

In the very early stages of star formation, during the prestellar core phase, SKA-Mid Band 5 spatially resolved observations will unveil the presence of complex carbon chains and rings, along with their formation and destruction mechanisms. This science is achievable with only a few hours of observation using the initial AA* configuration of SKA-Mid Band 5.

Observations targeting the protostellar envelopes of low-mass protostars, achieving high angular resolution ($\sim$ 500 au), will be performed with the SKA-Mid Band 5 AA4 configuration. These observations, requiring around 100 hours of integration time, will allow us to unveil the origin of the chemical dichotomy between hot corinos and WCCC sources, and to map and characterize accretion streamers. 

A SKA-Mid Band 5 key science project (1000 hours) focused on a typical star-forming region, such as OMC2, will enable the investigation of the chemistry of tens of protostellar objects down to the spatial scales of planet-forming disks. 

In high-mass star-forming regions, SKA-Mid Band 5 observations (hundreds of hours in AA4) will allow us to explore deuterated isotopologues and the chemistry of galactic environments at low metallicities.

The astrochemical community is deeply engaged in the preparatory work necessary to interpret and analyse these new data. This preparatory work includes conducting laboratory experiments and computational studies to provide molecular spectroscopic parameters, study chemical reactions, and revise the current astrochemical networks and models.
Notably, the SKAO will offer new opportunities of synergies with existing facilities such as ALMA. So far, the most powerful centimeter facility is the VLA in the northern hemisphere, which limits synergy with ALMA to the relatively small number of sources accessible from both observatories, leaving much of the southern sky unexplored. 

Finally, the current SKA-Mid design does not extend beyond 15 GHz, creating a critical observational gap between 15 GHz and the 30 GHz lower limit of ALMA. However, a potential upgrade of a frequency band for frequencies $>$ 15 GHz is technically possible (Band 6), as part of a future Observatory Development Programme. We highlight that this frequency range is essential for studies of chemical complexity, as several key molecular tracers fall within it, including ammonia (NH$_3$) and the strong methanol K-ladders around 25 GHz. In addition, many iCOMs can be more effectively traced in this range, where their transitions are brighter than in SKA-Mid Band 5 (\citealt{Jimenez2020, Ilee2020}; see also the \href{https://www.skao.int/sites/default/files/documents/d38-ScienceCase_band6_Feb2020.pdf}{SKA1 Beyond 15GHz:
The Science case for Band 6}. Access to these transitions provides unique insights into the history and chemical evolution of pre- and protostellar systems \citep[see e.g.,][]{DeSimone_2020ApJL, DeSimone2022ApJL}.\\
Furthermore, extending SKA-Mid to higher frequencies would contribute in expanding extragalactic studies. Indeed, the current extragalactic molecular studies  are only limited to a few bright or near galaxies, which prevent us from performing large chemical surveys, as it is the case with Galactic objects. In addition, even the highest currently planned frequency bands of SKA-Mid (Band 5a and 5b) have a too low sensitivity to perform extensive new molecular studies in nearby galaxies. Reaching  $\sim 50$ GHz with SKA-Mid would allow to target several important bright species, such as well-known temperature probes (NH$_3$ and H$_2$CO; e.g. \citealt{Ho&Townes1983, Mangum2013a, Mangum2013b}) whose transitions would fall near $\sim$ 25 GHz, 29 GHz, and 48 GHz. Other important dense gas tracers such as the cyanopolyynes HC$_3$N and HC$_5$N have several of their low-J lines falling in the range 18-50 GHz while at millimetre wavelengths, their emission is relatively faint in nearby galaxies. Enabling large studies of cyanopolyynes in nearby galaxies would allow to gain significant insight on the densest gas associated with star formation, as well as to establish how universal chemical complexity is in external galaxies.


\section*{Acknowledgements}
\textit{EB acknowledges the support from the Italian Ministry for Universities and Research under the Italian Science Fund (FIS 2 Call - Ministerial Decree No. 1236 of 1 August 2023) grant FIS-2023-00170 and the Next Generation EU funds within the National Recovery and Resilience Plan (PNRR), Mission 4 - Education and Research, Component 2 - From Research to Business (M4C2), Investment Line 3.1 - Strengthening and creation of Research Infrastructures, Project IR0000034 – “STILES - Strengthening the Italian Leadership in ELT and SKA". 
CeCe, ClCo, LP acknowledge the EC H2020 research and innovation
programme for: (i) the project "Astro-Chemical Origins” (ACO, No 811312), and (ii) the European Research Council (ERC) project “The Dawn of Organic Chemistry” (DOC, No 741002).
ClCo, LP, GS and ML acknowledge the PRIN-MUR 2020  BEYOND-2p (Astrochemistry beyond the second period elements, Prot. 2020AFB3FX), the project ASI-Astrobiologia 2023 MIGLIORA
(Modeling Chemical Complexity, F83C23000800005), the INAF-GO 2024 fundings ICES, and the INAF-GO 2023 fundings
PROTO-SKA (Exploiting ALMA data to study planet forming disks: preparing the advent of SKA, C13C23000770005). 
LP, ClCo, and GS also acknowledge financial support
under the National Recovery and Resilience Plan (NRRP), Mission 4, Component 2, Investment 1.1, Call for tender No. 104 published on 2.2.2022 by the Italian Ministry of University and Research (MUR), funded by the European Union – NextGenerationEU-Project Title 2022JC2Y93 Chemical Origins: linking the fossil composition of the Solar System with the chemistry of protoplanetary disks – CUP J53D23001600006 – Grant Assignment Decree No.
962 adopted on 30.06.2023 by the Italian Ministry of Ministry of University and Research (MUR). 
GS also acknowledges support from the INAF-Minigrant 2023 TRIESTE ("TRacing the chemIcal hEritage of our originS: from proTostars to planEts”). LC, IJ-S, and VMR, acknowledge support from grant no. PID2022-136814NB-I00 by MICIU/AEI/10.13039/501100011033 and by ERDF, UE. GE and PRM thank project PID-2022-137980NB-I00 funded by the Spanish Ministry of Science and Innovation / State Agency of Research MCIN / AEI/10.13039/501100011033 and by “ERDF A way of making Europe”.
LC acknowledges support from a fellowship from the ”la Caixa” Foundation (ID 100010434). The fellowship code is LCF/BQ/PR25/12110012.
VMR acknowledges support from the grant PID2022-136814NB-I00 by the Spanish Ministry of Science, Innovation and Universities/State Agency of Research MICIU/AEI/10.13039/501100011033 and by ERDF, UE;  the grant RYC2020-029387-I funded by MICIU/AEI/10.13039/501100011033 and by "ESF, Investing in your future", and from the Consejo Superior de Investigaciones Cient{\'i}ficas (CSIC) and the Centro de Astrobiolog{\'i}a (CAB) through the project 20225AT015 (Proyectos intramurales especiales del CSIC); and from the grant CNS2023-144464 funded by MICIU/AEI/10.13039/501100011033 and by “European Union NextGenerationEU/PRTR”. 
MP acknowledges the INAF grant 2023 MERCATOR (``MultiwavelEngth signatuRes of Cosmic rAys in sTar-fOrming Regions'')
and the INAF grant 2024 ENERGIA (``ExploriNg low-Energy cosmic Rays throuGh theoretical InvestigAtions at INAF''). MB acknowledges the support from the European Research Council (ERC) Advanced Grant MOPPEX 833460.
A~M~J acknowledges support from the Max Planck Society and SFB~1601. AC and PM received financial support from the European Research Council (ERC) under the European Union’s Horizon 2020 research and innovation programme (ERC Starting Grant “Chemtrip”, grant agreement No 949278). 
AS-M\ acknowledges support by: the grant PID2023-146675NB-I00 (MCI-AEI-FEDER, UE), the RyC2021-032892-I grant, and the Spanish program Unidad de Excelencia Mar\'ia de Maeztu CEX2020-001058-M, financed by MCIN/AEI/10.13039/501100011033, and by the MaX-CSIC Excellence Award MaX4-SOMMA-ICE.
PRM is a member of project PID2022-137980NB-I00, funded by MCIN/AEI/10.13039/501100011033/FEDER UE.
SW acknowledges support of the SNSF Eccellenza Professorial Fellowship PCEFP2\_181150. GB acknowledges support from the grant PID2023-146675NB-I00 (MCI-AEI-FEDER,
UE) and the grant CEX2024-001451-M funded by MICIU/AEI/10.13039/501100011033. The authors acknowledge the assistance of artificial intelligence in improving the readability of the
text.
}

\bibliographystyle{abbrvnat-maxbibnames4}
\bibliography{chapter} 

\end{document}